\documentstyle[times,psfig]{mn}

\newif\ifAMStwofonts
\ifoldfss

  \newcommand{\itl}[1] {{\it #1}}

  \ifCUPmtlplainloaded \else
    \NewTextAlphabet{textbfit} {cmbxti10} {}
    \NewTextAlphabet{textbfss} {cmssbx10} {}
    \NewMathAlphabet{mathbfit} {cmbxti10} {} % for math mode
    \NewMathAlphabet{mathbfss} {cmssbx10} {} %  "   "    "
  \fi
  \ifAMStwofonts
    \ifCUPmtlplainloaded \else
      \NewSymbolFont{upmath} {eurm10}
      \NewSymbolFont{AMSa} {msam10}
      \NewMathSymbol{\upi}     {0}{upmath}{19}
      \NewMathSymbol{\umu}     {0}{upmath}{16}
      \NewMathSymbol{\upartial}{0}{upmath}{40}
      \NewMathSymbol{\leqslant}{3}{AMSa}{36}
      \NewMathSymbol{\geqslant}{3}{AMSa}{3E}

      \let\leq=\leqslant \let\le=\leqslant
      \let\geq=\geqslant \let\ge=\geqslant
    \fi
  \fi
\fi % End of OFSS

\ifnfssone
  \newmathalphabet{\mathit}
  \addtoversion{normal}{\mathit}{cmr}{m}{it}
  \addtoversion{bold}{\mathit}{cmr}{bx}{it}

  \newcommand{\itl}[1] {\mathit{#1}}

  \newmathalphabet{\mathbfit} % math mode version of \textbfit{..}
  \addtoversion{normal}{\mathbfit}{cmr}{bx}{it}
  \addtoversion{bold}{\mathbfit}{cmr}{bx}{it}
  \newmathalphabet{\mathbfss} % math mode version of \textbfss{..}
  \addtoversion{normal}{\mathbfss}{cmss}{bx}{n}
  \addtoversion{bold}{\mathbfss}{cmss}{bx}{n}
  \ifAMStwofonts
    \ifCUPmtlplainloaded \else
      \UseAMStwoboldmath
      \makeatletter
      \new@mathgroup\upmath@group
      \define@mathgroup\mv@normal\upmath@group{eur}{m}{n}
      \define@mathgroup\mv@bold\upmath@group{eur}{b}{n}
      \edef\UPM{\hexnumber\upmath@group}
      \new@mathgroup\amsa@group
      \define@mathgroup\mv@normal\amsa@group{msa}{m}{n}
      \define@mathgroup\mv@bold\amsa@group{msa}{m}{n}
      \edef\AMSa{\hexnumber\amsa@group}
      \makeatother
      \mathchardef\upi="0\UPM19
      \mathchardef\umu="0\UPM16
      \mathchardef\upartial="0\UPM40
      \mathchardef\leqslant="3\AMSa36
      \mathchardef\geqslant="3\AMSa3E

      \let\leq=\leqslant \let\le=\leqslant
      \let\geq=\geqslant \let\ge=\geqslant
    \fi
  \fi
\fi % End of NFSS release 1

\ifnfsstwo

  \newcommand{\itl}[1] {\mathit{#1}}

  \DeclareMathAlphabet{\mathbfit}{OT1}{cmr}{bx}{it}
  \SetMathAlphabet\mathbfit{bold}{OT1}{cmr}{bx}{it}
  \DeclareMathAlphabet{\mathbfss}{OT1}{cmss}{bx}{n}
  \SetMathAlphabet\mathbfss{bold}{OT1}{cmss}{bx}{n}
  \ifAMStwofonts
    \ifCUPmtlplainloaded \else
      \DeclareSymbolFont{UPM}{U}{eur}{m}{n}
      \SetSymbolFont{UPM}{bold}{U}{eur}{b}{n}
      \DeclareSymbolFont{AMSa}{U}{msa}{m}{n}
      \DeclareMathSymbol{\upi}{0}{UPM}{"19}
      \DeclareMathSymbol{\umu}{0}{UPM}{"16}
      \DeclareMathSymbol{\upartial}{0}{UPM}{"40}
      \DeclareMathSymbol{\leqslant}{3}{AMSa}{"36}
      \DeclareMathSymbol{\geqslant}{3}{AMSa}{"3E}

      \let\leq=\leqslant \let\le=\leqslant
      \let\geq=\geqslant \let\ge=\geqslant
    \fi
  \fi
\fi % End of NFSS release 2

\ifCUPmtlplainloaded \else
  \ifAMStwofonts \else % If no AMS fonts
    \def\upi{\pi}
    \def\umu{\mu}
    \def\upartial{\partial}
  \fi
\fi

   \title[Formation and evolution of late-type dwarf galaxies. I]{Formation 
          and evolution of late-type dwarf galaxies. I. NGC\,1705 and 
          NGC\,1569}
   \author[D. Romano et al.]{Donatella Romano,$^{1}$\thanks{E-mail: 
           donatella.romano@bo.astro.it (DR); monica.tosi@bo.astro.it (MT); 
           matteucci@ts.astro.it (FM)} Monica Tosi,$^{1}$ and Francesca 
           Matteucci$^{2}$\\
           $^{1}$INAF\,--\,Osservatorio Astronomico di Bologna,
                 Via Ranzani 1, I-40127 Bologna, Italy\\
           $^{2}$Dipartimento di Astronomia, Universit\`a di Trieste,
                 Via G.B. Tiepolo 11, I-34131 Trieste, Italy}
   
\begin{document}

     \date{Accepted . Received ; in original form }

     \pagerange{\pageref{firstpage}--\pageref{lastpage}} \pubyear{2005}

     \maketitle

     \label{firstpage}

%%%%%%%%%%%%%%%%%%%%%%%%%%%%%%%%%%%%%%%%%%%%%%%%

   \begin{abstract}
     We present one-zone chemical evolution models for two dwarf starburst 
     galaxies, NGC\,1705 and NGC\,1569. Though especially designed for the 
     inner $\sim$~1 kpc region, where numerous H\,{\small II} regions and most 
     of the stars are observed, the models also account for the presence of 
     extended gaseous and dark matter haloes, and properly compute the binding 
     energy of the gas heated by supernova explosions. Using information about 
     the past star formation history and initial mass function of the systems 
     previously obtained from \textsl{Hubble Space Telescope} optical and 
     near-infrared colour-magnitude diagrams, we identify possible scenarios 
     of chemical enrichment and development of galactic winds. We assume that 
     the galactic winds are proportional to the Type II and Type Ia supernova 
     rates. As a consequence, they do not necessarily go to zero when the star 
     formation stops. In order not to overestimate the current metallicity of 
     the interstellar gas inferred from H\,{\small II} region spectroscopy, we 
     suggest that the winds efficiently remove from the galaxies the 
     metal-rich ejecta of dying stars. Conversely, requiring the final mass of 
     neutral gas to match the value inferred from 21-cm observations implies a 
     relatively low efficiency of interstellar medium entrainment in the 
     outflow, thus confirming previous findings that \emph{the winds driving 
     the evolution of typical starbursts are differential.} These conclusions 
     could be different only if the galaxies accrete huge fractions of 
     unprocessed gas at late times. By assuming standard stellar yields we 
     obtain a good fit to the observed nitrogen to oxygen ratio of NGC\,1569, 
     while the mean N/O ratio in NGC\,1705 is overestimated by the models. 
     Reducing the extent of hot bottom burning in low-metallicity 
     intermediate-mass stars does not suffice to solve the problem. Localized 
     self-pollution from stars more massive than 60~M$_\odot$ in NGC\,1705 
     and/or funneling of larger fractions of nitrogen through its winds are 
     then left to explain the discrepancy between model predictions and 
     observations. Inspection of the log(N/O) vs. log(O/H)+12 diagram for a 
     large sample of dwarf irregular and blue compact dwarf galaxies in the 
     literature favours the latter hypothesis. Yet, the physical mechanisms 
     responsible for such a selective loss of metals remain unclear.
   \end{abstract}


%%%%%%%%%%%%%%%%%%%%%%%%%%%%%%%%%%%%%%%%%%%%%%%%

   \section{Introduction}

   Low-luminosity galaxies are the most common type of galaxies in the nearby 
   universe (Hodge 1971). Dwarf galaxies present a well pronounced dichotomy 
   between the two main classes of early-type and late-type dwarfs (dwarf 
   irregulars, DIGs, and blue compact dwarfs, BCDs). Indeed, most dwarfs are 
   either gas-poor spheroids with dominant intermediate-age or old stellar 
   populations, or gas-rich, star-forming systems (van den Bergh 1977). Only 
   few transition objects are known (Sandage \& Hoffman 1991). Here we 
   concentrate on DIGs and BCDs. BCDs are currently undergoing vigorous star 
   formation (SF) activity. They are objects of compact appearance, with 
   centrally concentrated starburst, gas and star distributions. Part of the 
   extended neutral gas may be kinematically decoupled from the galaxies (van 
   Zee, Skillman \& Salzer 1998). DIGs are instead dominated by scattered 
   bright H\,{\small II} regions in the optical, while in H\,{\small I} they 
   show a complicated fractal-like pattern of shells, filaments and clumps. 
   Typical H\,{\small I} masses are $\leq$ 10$^9$ M$_\odot$.

   Studying gas-rich starbursting objects helps to better understand the 
   process of star formation from low-metallicity gas clouds and the 
   derivation of the primordial helium abundance with the need for a minimum 
   extrapolation to early conditions (e.g., Peimbert \& Torres-Peimbert 1974; 
   Izotov \& Thuan 1998). Moreover, by studying dwarf galaxies, one can 
   complement the information available for solar composition stars and get 
   useful insights into the metal dependence of the stellar yields and the 
   winds in Wolf-Rayet (W-R) stars. 

   Late-type dwarfs have low metallicities, large gas content and mostly young 
   stellar populations: all these features indicate they are poorly evolved 
   objects and are consistent with two possible explanations for their nature: 
   (i) either they are newly formed galaxies or (ii) they have evolved slowly 
   over the Hubble time. In their pioneering work, Searle, Sargent \& Bagnuolo 
   (1973) discarded the youth hypothesis and concluded that extremely blue 
   galaxies must have undergone brief but intense bursts of SF separated by 
   long quiescent periods (bursting SF). Recent detections of old underlying 
   stellar populations in most BCDs seem to corroborate Searle's et al. (1973) 
   earlier suggestion and reveal at least another burst of SF besides the 
   present one, even in the case of the most metal-poor BCD known, I\,Zw\,18 
   (Aloisi, Tosi \& Greggio 1999; \"Ostlin 2000; Izotov \& Thuan 2004).

   Besides the bursting SF mode, gasping (Tosi et al. 1991) or mild continuous 
   (Carigi, Col\'\i n \& Peimbert 1999; Legrand et al. 2000) SF regimes have 
   been proposed for DIGs and BCDs. The gasping regime, in which the 
   interburst periods last significantly less than the active phases, probably 
   is the more realistic picture for many of them (Schulte-Ladbeck et al. 
   2001). 

   Larson \& Tinsley (1978) postulated the existence of a close connection 
   between violent dynamical phenomena and bursts of SF. Observational 
   evidence of intense ongoing SF in isolated BCDs (Campos-Aguilar \& Moles 
   1991) contradicts this hypothesis. However, at least some `isolated' BCDs 
   might have low surface brightness or pure H\,{\small I} companions 
   (M\'endez \& Esteban 2000; Noeske et al. 2001; Pustilnik et al. 2001; 
   Hoffman et al. 2003). In addition to tidal interactions, low-mass stellar 
   or gaseous companions might trigger and fuel starbursts by infalling on to 
   the DIG (K\"oppen \& Edmunds 1999; cfr. also the `minor merger' phenomenon 
   described by Sancisi 1999). Another explanation for the intermittent mode 
   of SF in DIGs and BCDs was given by Gerola, Seiden \& Schulman (1980), who 
   applyed the stochastic self-propagating SF (SSPSF) model to systems of 
   small size. Yet, the high dispersion of the properties of these galaxies 
   cannot be explained in terms of the SSPSF mechanism alone (Matteucci \& 
   Chiosi 1983). In particular, galactic winds (GWs) of variable intensity 
   (Matteucci \& Chiosi 1983; Matteucci \& Tosi 1985) releasing metals in 
   different proportions (enriched winds -- Pilyugin 1993; Marconi, Matteucci 
   \& Tosi 1994; Carigi et al. 1995) were introduced in order to reproduce the 
   high $\Delta Y$/$\Delta$O ratio and the distribution of N/O vs. O/H 
   observed in extragalactic H\,{\small II} regions.

   Observational evidence for starburst-driven large-scale outflows was 
   established for a number of starburst or young post-starburst systems, 
   including I\,Zw\,18 (Martin 1996), NGC\,1569 (Waller 1991; Heckman et al. 
   1995), and NGC\,1705 (Meurer et al. 1992; Marlowe et al. 1995). This 
   evidence is now growing fast, since diffuse and filamentary gas associated 
   with large-scale outflows are detected for an increasing number of 
   starbursting dwarfs (Veilleux et al. 2003; Cannon et al. 2004). However, 
   whether the final fate of this gas is to eventually escape from the galaxy 
   or to cool and recollapse forming the next generations of stars is not yet 
   clear. \textsl{Far Ultraviolet Spectroscopic Explorer (FUSE)} observations 
   of the O\,{\small VI} $\lambda$\,1032 absorption line arising in the 
   galactic outflow of NGC\,1705 suggest that the superbubble has begun to 
   blow out of the interstellar medium (ISM) in the disk of NGC\,1705 (Heckman 
   et al. 2001). This superbubble is likely to become a superwind and vent its 
   metals and kinetic energy. The bulk of the global ISM (H\,{\small I}) in 
   NGC\,1705 will be retained (Mac Low \& Ferrara 1999). A similar scenario 
   should be suitable also for I\,Zw\,18 (Martin 1996). Recently, we have also 
   been provided with the first direct evidence for metal-enriched winds from 
   dwarf starburst galaxies: deep \textsl{Chandra} imaging of NGC\,1569 has 
   led to an estimated mass of oxygen in the hot wind similar to the oxygen 
   yield of the current starburst (Martin, Kobulnicky \& Heckman 2002). The 
   most likely interpretation is that the wind carries nearly all the metals 
   expelled by supernovae (SNe). A similar scenario emerges also from 
   chemodynamical simulations (e.g. Recchi et al. 2004, 2005).

   In this paper we study the chemical evolution of DIGs and BCDs. In 
   particular, we compare predictions from models with structural parameters 
   applicable to NGC\,1705 and NGC\,1569 with their observed properties. Our 
   choice is motivated by the fact that the recent star formation history 
   (SFH), initial mass function (IMF) and chemical enrichment of these two 
   galaxies are rather well constrained by the observations. Therefore, we can 
   construct models without making arbitrary assumptions about the SFH and IMF 
   (which are usually free parameters of the chemical evolution models), and 
   assume the values inferred from the observations. As a consequence, we are 
   able to better constrain other poorly known physical processes which drive 
   galaxy evolution, such as, for instance, the stellar feedback. Both 
   NGC\,1705 and NGC\,1569 have strong galactic winds, likely triggered by the 
   recent starbursts, and we aim at better understanding how these outflows 
   develop.

   The layout of the paper is the following. Observations are presented in 
   \S~2, the models are described in \S~3, model results are given in \S~4 and 
   discussed in \S~5. Finally, the conclusions are drawn in \S~6.

   \section{Observations}

   \subsection{General properties of dwarf irregular and blue compact dwarf 
               galaxies}

   \subsubsection{Chemical abundances} \label{secObsChem}

   Metallicities in DIGs and BCDs are usually derived from the ionized gas in 
   H\,{\small II} regions. From the optical, O, N, S, Ne, Ar, He, and even 
   fainter lines such as those of Fe are measured, while C and Si abundances 
   are obtained from the ultraviolet (UV). Oxygen is the element with the most 
   reliable determinations, since all its most important ionization stages can 
   be observed. The intrinsic uncertainty is of the order of $\sim$ 0.1 dex, 
   which rises at 0.2 dex or more when the electron temperature cannot be 
   directly determined (Pagel 1997). H\,{\small II} regions are ionized by 
   newly born massive stars, hence showing the metallicity of the ISM at the 
   present time. A possible problem with metal-poor H\,{\small II} regions 
   could be self-pollution of fresh metals by winds from young massive stars, 
   so that the abundances inferred from the nebular emission lines would not 
   be representative of that in the local ISM (Kunth \& Sargent 1986; Pagel, 
   Terlevich \& Melnick 1986). However, if the elements ejected from massive 
   stars are in high ionization stages, they are not expected to significantly 
   contribute to the element abundances as derived from optical lines 
   (Pantelaki \& Clayton 1987). Observational evidence for this comes from the 
   chemical homogeneity of most star-forming galaxies, in which the expected 
   spatial variations in oxygen abundances are not observed, despite the 
   presence of multiple massive star clusters (Kobulnicky \& Skillman 1997, 
   1998).

   Recently, a step forward in understanding the physical and chemical 
   evolution of dwarf star-forming galaxies has been accomplished thanks to 
   \textsl{FUSE}, which makes possible the determination of the metal 
   abundances of the neutral ISM of these galaxies. Preliminary data analysis 
   points to an offset in metal content between the neutral ISM and H\,{\small 
   II} regions, at least for the galaxies surveyed up to now (Aloisi et al. 
   2005). We naturally find a qualitative explanation for this offset in the 
   framework of our model (see discussion in Sect.~4). However, many 
   uncertainties still affect both the interpretation of the data and the 
   physics of the model, thus making the accurate quantification of the offset 
   a hardly achievable goal.

   Only for the closest galaxies, there is the possibility of tracing the 
   history of the chemical enrichment of the ISM through spectroscopy of the 
   resolved stellar populations (Hill 2004; Tolstoy et al. 2004, and 
   references therein). Photometric estimates of the iron stellar abundances 
   can be obtained from colour-magnitude diagrams (CMDs), but with much lower 
   reliability.

   \subsubsection{Star formation history and initial mass function}

   The synthetic CMD method (Tosi et al. 1991; see also Gallart et al. 1994; 
   Tolstoy \& Saha 1996; Aparicio \& Gallart 2004 and references therein) 
   allows one to estimate the epochs and intensities of the SF activity in a 
   galaxy, as well as the IMF slope.

   In galaxies outside the Local Group, crowding and magnitude limits make it 
   increasingly difficult to resolve the fainter stars and shorten accordingly 
   the reachable look-back time. Nevertheless, their SFHs have been inferred 
   up to a few Gyr ago (Tosi 2003, and references therein).

   The following sections, Sects.~\ref{sec1705} and \ref{sec1569}, are mostly 
   devoted to a detailed description of the SFHs, IMF slopes, and chemical 
   abundances inferred for NGC\,1705 and NGC\,1569. In spite of being 
   differently classified (as a BCD NGC\,1705 and as a DIG NGC\,1569), they 
   are otherwise similar objects, showing both evidence of a complex structure 
   and the presence of galactic outflows.
   
   \subsection{Individual objects} \label{secStarFor}

   \subsubsection{NGC\,1705} \label{sec1705}

%%%%%%%%%%%%%%%%%%%%%%%%%%%%%%%%%%%%%%%%%%%%%%%%
%
   \begin{table}
   \caption{NGC\,1705 identity card.}
   \label{tab1705}
   \begin{tabular}{@{}lcll@{}}
   \hline
   Quantity & \multicolumn{2}{c}{Observed value} & References \\
   \hline
   $D$                  & \multicolumn{2}{c}{$5.1 \pm 0.6$ Mpc}                
               & 1 \\
   $R_{\mathrm{H}}$     & \multicolumn{2}{c}{$1.7$ kpc$^{\mathrm{a}}$}         
               & 2 \\
   $M_{\mathrm{gas}}$   & \multicolumn{2}{c}{$1.7 \times 10^8$ 
M$_\odot^{\mathrm{\;\;\;a}}$}    & 3 \\
   $M_{\mathrm{stars}}$ & \multicolumn{2}{c}{$1.7 \times 10^8$ 
M$_\odot^{\mathrm{\;\;\;a}}$}    & 3 \\
   $Z$                  & \multicolumn{2}{c}{$0.004$}                          
               & 4 \\
                        & H\,{\tiny II} regions                             & 
       Neutral ISM                  & \\
   $\log$(O/H) + 12     & 8.46                                              &
                                    & 3 \\
                        & 8.36                                              &
                                    & 4 \\
                        & 8.0                                               &
                                    & 5 \\
                        & $8.21 \pm 0.05$                                   & 
       $7.50 \pm 0.01^{\mathrm{b}}$ & 6, 7 \\
   $\log$(N/H) + 12     & $6.46 \pm 0.08$                                   & 
       $5.68 \pm 0.06$              & 6, 7 \\
   $\log$(Fe/H) + 12    &                                                   & 
       $6.21 \pm 0.03$              & 7 \\
   $\log$(N/O)          & $-1.75 \pm 0.06^{\mathrm{c}}$                     & 
       $-1.82 \pm 0.07$             & 6, 7 \\
                        & $-1.62$                                           &
                                    & 4 \\
   $\log$(Ne/O)         & $-0.426 \pm 0.033$                                & 
                                    & 6 \\
   \hline
   \end{tabular}

   \medskip
   1-- Tosi et al. 2001; 2-- Meurer et al. 1998; 3-- Meurer et al. 1992; 4-- 
   Storchi-Bergmann et al. 1994; 5-- Heckman et al. 1998; 6-- Lee \& Skillman 
   2004; 7-- Aloisi et al. 2005.
   \begin{list}{}{}
   \item[$^{\mathrm{a}}$] Radius and masses were modified to reflect the 
                          distance used here.
   \item[$^{\mathrm{b}}$] The oxygen abundance is that derived from the total 
                          (neutral plus ionized) absorbing gas. That referring 
                          to the sole neutral gas would be somehow lower.
   \item[$^{\mathrm{c}}$] $\log$(N/O) = $-$1.63 $\pm$ 0.07 if the anomalously 
                          low nitrogen value in region B4 is ignored (see Lee 
                          \& Skillman 2004).
   \end{list}
   \end{table}
%
%%%%%%%%%%%%%%%%%%%%%%%%%%%%%%%%%%%%%%%%%%%%%%%%

%%%%%%%%%%%%%%%%%%%%%%%%%%%%%%%%%%%%%%%%%%%%%%%%
%
   \begin{table}
   \caption{Average SFRs at various epochs for NGC\,1705 as inferred from 
            the observations (Annibali et al. 2003).}
   \label{tab1705sfr}
   \begin{tabular}{@{}cc@{}}
   \hline
   SFR (M$_\odot$ yr$^{-1}$) & Look-back time (Myr) \\
   \hline
   $5.6 \times 10^{-2}$ & $1000$--$5000$ \\
   $5.8 \times 10^{-2}$ & $50$--$1000$   \\
   $7.7 \times 10^{-3}$ & $15$--$50$     \\
   $6.8 \times 10^{-2}$ & $10$--$15$     \\
   $0.314$              & $0$--$3$       \\
   \hline
   \end{tabular}
   \end{table}
%
%%%%%%%%%%%%%%%%%%%%%%%%%%%%%%%%%%%%%%%%%%%%%%%%

   NGC\,1705 is a nearby, isolated BCD. Its nuclear region contains an 
   extremely luminous super star cluster (SSC), responsible for about half of 
   the total UV light from the galaxy (Meurer et al. 1992). SF is found to 
   occur now, mainly concentrated in the high surface brightness part of the 
   galaxy (Annibali et al. 2003). In this region, the presence of W-R emission 
   lines suggests ongoing SF during the past 5 Myr (Meurer et al. 1992). In 
   H$\alpha$, NGC\,1705 shows a bipolar morphology, suggesting an outflow 
   powered by the recent SF activity in the nucleus (Meurer et al. 1992). UV 
   absorption-line kinematics bears witness to this too (Heckman \& Leitherer 
   1997). The gas metallicity, as derived from UV, optical and near infrared 
   (NIR) spectra of H\,{\small II} regions is 12+log(O/H) = 8.36 
   (Storchi-Bergmann, Calzetti \& Kinney 1994), corresponding to $Z \simeq$ 
   0.004 (for $Z_\odot$ = 0.017), close to the metallicity of the SMC. A 
   higher value, 12+log(O/H) $\sim$ 8.46, is reported by Meurer et al. (1992), 
   while Heckman et al. (1998) give 12+log(O/H) $\simeq$ 8.0. Recently, Lee \& 
   Skillman (2004) have derived a mean oxygen abundance of 12+log(O/H) = 8.21 
   $\pm$ 0.05. They have detected [O\,{\small III}] $\lambda$\,4363 in five 
   H\,{\small II} regions of NGC\,1705, which makes their oxygen determination 
   highly reliable, thanks to the direct measurements of electron 
   temperatures. The estimated log(N/O) = $-$1.75 $\pm$ 0.06 is among the 
   lowest values ever observed for dwarf irregulars, but it raises up to 
   $-$1.63 $\pm$ 0.07 if the anomalously low nitrogen abundance in H\,{\small 
   II} region B4 is ignored\footnote{Notice that a value of log(N/O) = $-$1.63 
   $\pm$ 0.07 is in excellent agreement with the previous estimate by 
   Storchi-Bergmann et al. (1994) -- see Table~\ref{tab1705}, second-last 
   row.}. In any case, we caution that nitrogen could not be determined in the 
   same regions where direct electron temperatures were available. 
   \textsl{FUSE} spectra seem to suggest that the N/O ratio in H\,{\small I} 
   is pretty much the same, while the oxygen content is lower in the cold gas 
   (Aloisi et al. 2005).

   Deep and accurate photometry has allowed to resolve stars from the most 
   central regions to the extreme outskirts (Tosi et al. 2001), so that it has 
   been possible to divide the galaxy in several roughly concentric regions 
   and to derive the SFH for each of them (Annibali et al. 2003): (i) 
   different zones of the galaxy experienced different SFHs; (ii) the derived 
   SF is almost continuous with fluctuations in the SFR (i.e., gasping rather 
   than bursting); (iii) the young population dominates in the inner regions, 
   where a recent, intense burst occurred; (iv) the strength of the most 
   recent activity gradually decreases moving from the central to the outer 
   regions; (v) NGC\,1705 is definitely not a young galaxy: an old population 
   is spread over the whole galaxy (see also Meurer et al. 1992); (vi) the 
   total stellar mass astrated in the galaxy is $\sim$~3~$\times$~10$^8$~ 
   M$_\odot$ and $\sim$~20 per cent of the stars are younger than 1 Gyr; (vii) 
   a Salpeter ($x$~= 1.35) or slightly steeper ($x$~= 1.6) field star IMF 
   provides the best agreement with the observations.

   In Table~\ref{tab1705} we summarize some structural parameters and observed 
   properties of NGC\,1705. Listed are: the distance to NGC\,1705, $D$; its 
   Holmberg radius, $R_{\mathrm{H}}$; its total atomic gas mass (H\,{\small I} 
   + He), $M_{\mathrm{gas}}$; its stellar mass, $M_{\mathrm{stars}}$; its 
   present-day metallicity, $Z$, and the mean oxygen, nitrogen, iron and neon 
   abundances measured in H\,{\small II} regions and neutral gas. 
   Table~\ref{tab1705sfr} reports the average SFRs in representative age bins 
   for the whole surveyed area, $\sim$ 12.5 kpc$^2$, as listed in Annibali et 
   al. (2003 -- their table 6, last row).

   \subsubsection{NGC\,1569} \label{sec1569}

   NGC\,1569 is an exceptionally active DIG, which contains three SSCs (De 
   Marchi et al. 1997; Origlia et al. 2001; Sirianni et al. 2005) and many 
   H\,{\small II} regions out of the SSC zone, where the SF is active now 
   (Waller 1991). According to Aloisi et al. (2001), young stars ($m >$ 8 
   M$_\odot$, $\tau_m <$ 50 Myr) are mostly clustered around the three SSCs, 
   whereas intermediate-age objects (1.9 $< m/$M$_\odot <$ 8, 50~Myr~$< \tau_m 
   <$~1~Gyr) are more evenly distributed. Older stars ($m <$ 1.9 M$_\odot$, 
   $\tau_m >$ 1 Gyr) are visible at the outskirts of the starbursting region. 
   Filaments are found at different wavelengths and a strong spatial 
   correlation is present between the extended X-ray emission and the 
   H$\alpha$ filaments detected in the optical (Martin et al. 2002). In 
   H\,{\small I}, a dense, clumpy ridge distribution appears, surrounded by 
   more extended, diffuse neutral hydrogen. Discrete features such as arms and 
   bridges are observed (Stil \& Israel 2002) and the dust properties 
   (Galliano et al. 2003) are consistent with the presence of shocks. Perhaps, 
   the galaxy is presently ingesting a companion gaseous cloud (Stil \& Israel 
   1998; M\"uhle et al. 2005), and it is tempting to speculate upon the 
   enhanced SF to coincide with the accretion episode. Recent SN explosions 
   triggered a GW (Martin et al. 2002), but it is unlikely that it will 
   efficiently remove the ISM, unless stripping or other environmental factors 
   intervene (D'Ercole \& Brighenti 1999; see also Mac Low \& Ferrara 1999). 
   \textsl{Chandra} observations reveal large inhomogeneities in the metal 
   abundance of the ISM, with $Z$ ranging from 0.1 to 1 $Z_\odot$ (Martin et 
   al. 2002), again suggesting that multiple SN explosions from recent 
   starbursts polluted the ISM.

%%%%%%%%%%%%%%%%%%%%%%%%%%%%%%%%%%%%%%%%%%%%%%%%
%
   \begin{table}
   \caption{NGC\,1569 identity card.}
   \label{tab1569}
   \begin{tabular}{@{}lcl@{}}
   \hline
   Quantity & Observed value & References \\
   \hline
   $D$                & $2.2 \pm 0.6$ Mpc$^{\mathrm{a}}$      & 1 \\
   $d$                & $1.85$ kpc                            & 2 \\
   $M_{\mathrm{gas}}$ & $(1.5 \pm 0.3) \times 10^8$ M$_\odot$ & 1 \\
   $M_{\mathrm{tot}}$ & $3.3 \times 10^8$ M$_\odot$           & 1 \\
   $\mu$              & $0.46 \pm 0.09$                       & 1 \\
   ${ }$              & $0.264$--$0.73$                       & 3, 4 \\
   $Z$                & $0.004$                               & 5 \\
   (He/H)             & $0.080 \pm 0.003^{\mathrm{b}}$        & 6 \\
   $\log$(O/H) + 12   & $8.19$--$8.37$                        & 6, 7, 8 \\
   $\log$(N/O)        & $-1.39 \pm 0.05$                      & 6 \\
                      & $-0.81$                               & 8 \\
   SFR                & $\sim 0.5$ M$_\odot$ yr$^{-1 \mathrm{c}}$        
                                                              & 9 \\
                      & $0.32$ M$_\odot$ yr$^{-1}$            & 10 \\
   \hline
   \end{tabular}

   \medskip
   1-- Israel 1988; 2-- Stil \& Israel 2002; 3-- Lee et al. 2003a; 4-- Martin 
   et al. 2002; 5-- Gonz\'alez Delgado et al. 1997; 6-- Kobulnicky \& Skillman 
   1997; 7-- Martin 1997; 8-- Storchi-Bergmann et al. 1994; 9-- Greggio et al. 
   1998; 10-- Hunter \& Elmegreen 2004.
   \begin{list}{}{}
   \item[$^{\mathrm{a}}$] Recently, Makarova \& Karachentsev (2003) have 
                          derived from the I magnitude of the tip of the RGB 
                          two possible values for the distance to NGC\,1569, 
                          1.95 $\pm$ 0.2 Mpc and 2.8 $\pm$ 0.2 Mpc.
   \item[$^{\mathrm{b}}$] The error bar is likely to be larger (see Olive \& 
                          Skillman 2004 reanalysis of helium abundance 
                          determinations in extragalactic H\,{\tiny II} 
                          regions).
   \item[$^{\mathrm{c}}$] The tabulated value refers to a look-back time of 
                          0.1--0.15 Gyr. It has been derived for the central 
                          $\sim$ 0.14 kpc$^2$ of the galaxy, assuming a 
                          Salpeter IMF. The data also indicate a halt in the 
                          SFR 5--10 Myr ago. Strong SF activity is expected 
                          between 1.5 and 0.15 Gyr ago.
   \end{list}
   \end{table}
%
%%%%%%%%%%%%%%%%%%%%%%%%%%%%%%%%%%%%%%%%%%%%%%%%

   The SFH of NGC\,1569 has been extensively studied in the last decade. 
   Vallenari \& Bomans (1996), from V and I optical bands, find that a global 
   SF episode occurred from 100 to 4 Myr ago. They find hints for an older 
   episode (from 1.5 Gyr to 150 Myr ago) and rule out the existence of 
   long-lasting quiescent phases in the last 1.5 Gyr. Similarly, based on B 
   and V bands, Greggio et al. (1998) derive for the most recent burst -- 
   which, in the examined region, ended from 5 to 10 Myr ago -- a duration of 
   $\geq$ 100 Myr and a very high rate, $\sim$ 0.5 M$_\odot$ yr$^{-1}$ in the 
   inner $\sim$ 0.14 kpc$^2$ of the galaxy if Salpeter's IMF is assumed over 
   the 0.1--120 M$_\odot$ stellar mass range. If quiescent periods occurred, 
   they lasted less than $\sim$ 10 Myr. By using NIR data, Angeretti et al. 
   (2005) looked deeper backward in time and found three major episodes of SF 
   in the last 1 (possibly 2) Gyr. They are able to distinguish two episodes 
   at least for the most recent SF activity. The first started $\sim$ 150 Myr 
   ago and ended $\sim$ 40 Myr ago. The second formed stars between 37 and 13 
   Myr ago, at a mean rate about 3 times higher than the preceding one, i.e. 
   $\sim$ 0.13 M$_\odot$ yr$^{-1}$ if Salpeter's IMF is assumed, in agreement 
   with previous literature data. From their \textsl{Hubble Space Telescope 
   (HST)} NICMOS/NIC2 data, the same authors also find evidence for a gap in 
   the SF activity (or, alternatively, for a low-level SF) from 300 to 150 Myr 
   ago, and traces of an older SF episode, occurred about 1--2 Gyr ago at a 
   rate between 0.01 and 0.06 M$_\odot$ yr$^{-1}$ (depending on the assumed 
   duration). The relatively recent, strong SF activity proceeded at rates 
   that could have not been sustained over a Hubble time, since it would have 
   consumed all the available gas in $\sim$ 1 Gyr, or even less (Greggio et 
   al. 1998). Although a low-level SF is not excluded in the past, certainly 
   the last 1--2 Gyr have been peculiar in the SFH of NGC\,1569. There is an 
   overall agreement among the various SFHs inferred from different bands.

   In a recent work, Anders et al. (2004) from multi-colour archive 
   \textsl{HST} data show that a high level of star cluster formation started 
   $\sim$ 25 Myr ago in NGC\,1569, while a lower level of activity 
   characterized older epochs, with a secondary peak $\sim$ 100 Myr ago. The 
   age distribution of star clusters displays a gap in the age interval 
   160--400 Myr. Heckman et al. (1995) find that observations of the 
   expanding superbubbles in NGC\,1569 can be fairly well explained assuming a 
   SF episode at a constant rate over the ages 13--32 Myr, which are very 
   similar to those inferred by Angeretti et al. (2005) for their youngest 
   burst. Hunter \& Elmegreen (2004) find a present-day SFR 
   $\psi(t_{\mathrm{now}}) \simeq$ 0.32 M$_\odot$ yr$^{-1}$ from H$\alpha$ 
   imaging.

   For reference, in Table~\ref{tab1569} we summarize some basic observational 
   properties of NGC\,1569. Listed are: the distance to NGC\,1569, $D$; its 
   maximum optical size, $d$; its total atomic gas mass (H\,{\small I} + He), 
   $M_{\mathrm{gas}}$; its total mass, $M_{\mathrm{tot}}$; the ratio between 
   the latter two, $\mu$, which shows that from 1/4 to 3/4 of the total mass 
   eventually resides in neutral atomic gas, depending on different authors; 
   the total metal, helium, oxygen and nitrogen abundances in the ISM of 
   NGC\,1569 and, finally, a lower limit to the intensity of the latest SF 
   episode, according to optical data. It is worth noticing that the N/O ratio 
   measured by Kobulnicky \& Skillman (1997) for NGC\,1569 is significantly 
   lower than that derived by Storchi-Bergmann et al. (1994). We will compare 
   only the more recent, more accurate determinations to our model predictions.

%%%%%%%%%%%%%%%%%%%%%%%%%%%%%%%%%%%%%%%%%%%%%%%%
%
% Holmberg radius: radius at which the surface brightness is 
%                  26.5 mag arcsec-2 in B band, good measure of the 
%                  extent of the bulk of the stars in a galaxy.
%
% Effective radius: distance from the center of a galaxy from within 
%                   which half the galaxy's luminosity is emitted.
%
%%%%%%%%%%%%%%%%%%%%%%%%%%%%%%%%%%%%%%%%%%%%%%%%

   \section{Models}

   In order to study the formation and evolution of late-type dwarfs, we use 
   an updated version of the galactic chemical evolution (GCE) model developed 
   by Bradamante, Matteucci \& D'Ercole (1998). In particular: (i) we adopt 
   up-to-date stellar yields, (ii) we revise the prescriptions about the GWs, 
   and (iii) we treat the infall of primordial gas as a free parameter. The 
   adopted SFH and IMF are those inferred from previous CMD analyses, less 
   prone to \emph{ad hoc} assumptions.

   The main features of the model are the following:
   \begin{enumerate}
     \item the model is one zone with instantaneous and complete mixing of gas 
           inside it;
       \item the instantaneous recycling approximation is relaxed, i.e. the 
             stellar lifetimes are taken into account in detail;
         \item GWs originate when the thermal energy of the gas equates its 
               binding energy;
	   \item the winds preferably expel metals from the galaxy, while 
	         retaining large fractions of hydrogen;
	     \item the GW efficiency varies with time according to the SN (II 
		   plus Ia) rate predicted by the model.
   \end{enumerate}

   \subsection{General prescriptions} \label{secModels}

   Homogeneity and instantaneous mixing of gas are reasonable working 
   hypotheses, as long as gradients are not observed in the galaxy and the 
   cooling time of the gas proves to be short enough. Hydrodynamical 
   simulations of the behaviour of the ISM and the metals ejected by massive 
   stars after a starburst suggest that most of the metals effectively cool 
   off in a few million years (Hensler et al. 2004; Recchi et al. 
   2001)\footnote{Notice that these results strongly rest on the assumed (low) 
   heating efficiency of SNeII, i.e. on the (low) fraction of the SN explosion 
   energy which remains effectively stored in the ISM and is not radiated 
   away, $\eta_{\mathrm{SNII}}$ = 0.03. The larger $\eta_{\mathrm{SNII}}$, the 
   longer the cooling time of the ejecta.}. The existence of a small age gap 
   between previous SF activity and the latest ongoing burst in NGC\,1705 
   ($\sim$ 6 million years; see Table~\ref{tab1705sfr}) is consistent with 
   quite short time scales for gas cooling as well. Yet, the low occurrence of 
   localized chemical pollution in the vicinity of young star clusters in 
   star-forming dwarfs (see Sect.~\ref{secObsChem}) points to time scales for 
   metals cooling longer than $\sim$ 10$^7$ yr (Kobulnicky \& Skillman 1997). 

%%%%%%%%%%%%%%%%%%%%%%%%%%%%%%%%%%%%%%%%%%%%%%%%
%
   \begin{table*}
   \begin{minipage}{17.5cm}
   \caption{Parameters of the models with their adopted values.}
   \label{tabcommonpar}
   \begin{tabular}{@{}lp{8cm}cc@{}}
   \hline
   Parameter & Meaning & \multicolumn{2}{c}{Adopted value} \\
             &         &       NGC\,1705 & NGC\,1569       \\
   \hline
   $M_{\mathrm{nor}}$                & 
     normalizing mass (gaseous matter available for accretion inside the DM 
     potential) & 7 $\times$ 10$^8$ M$_\odot$ & 5 $\times$ 10$^8$ M$_\odot$ \\ 
   $M_{\mathrm{d}}$                  & 
     DM amount & 7 $\times$ 10$^9$ M$_\odot$ & 5 $\times$ 10$^9$ M$_\odot$ \\
   $R_{\mathrm{eff}}$                & 
     effective radius of the visible matter & 1.2 kpc & 1 kpc \\
   $R_{\mathrm{eff}}/R_{\mathrm{d}}$ & 
     effective to DM core radii ratio & \multicolumn{2}{c}{0.1} \\
   $\tau$                            & 
     time scale for mass accretion & 8 Gyr & 0.5, 5 Gyr \\
   $\eta_{\mathrm{SNII}}$            & 
     SNII thermalization efficiency & 0.20 & 0.03, 0.20, 0.50 \\
   $\eta_{\mathrm{SNIa}}$            & 
     SNIa thermalization efficiency & 0.20 & 0.20, 0.50, 1.00 \\
   $\eta_{\mathrm{wind}}$            & 
     stellar wind thermalization efficiency & \multicolumn{2}{c}{0.20} \\
   $t_{\mathrm{now}}$                & 
     age of the universe & \multicolumn{2}{c}{13.7 Gyr} \\
   \hline
   \end{tabular}
   \end{minipage}
   \end{table*}
%
%%%%%%%%%%%%%%%%%%%%%%%%%%%%%%%%%%%%%%%%%%%%%%%%

   Galactic outflows develop when the thermal energy of the gas, 
   $E_{\mathrm{th}}$, equates its binding energy, $E_{\mathrm{b}}$. Following 
   Bradamante et al. (1998), the thermal energy of the gas grows as a 
   consequence of SF activity due to energy injection from stellar winds and 
   SN explosions (both Type II and Type Ia SNe are considered):
   {\setlength\arraycolsep{2pt}
    \begin{eqnarray}
     E_{\mathrm{th}}(t) & = & \eta_{\mathrm{SNII}} E_{\mathrm{SN}} \int_0^t 
                            R_{\mathrm{SNII}}(t') \ {\mathrm{d}}t' \,\, +{}
                            \nonumber\\
                        & & {}+ \eta_{\mathrm{SNIa}} E_{\mathrm{SN}} \int_0^t 
                            R_{\mathrm{SNIa}}(t') \ {\mathrm{d}}t' \,\, +{}
                            \nonumber\\
                        & & {}+ \eta_{\mathrm{wind}} E_{\mathrm{wind}} 
			    \int_0^t\!\!\!\int_{12}^{100} \varphi(m) \psi(t') \
			    {\mathrm{d}}m \ {\mathrm{d}}t'{}.
    \end{eqnarray}}
   $\varphi(m)$ is the IMF, $\psi(t)$ is the SFR, $E_{\mathrm{SN}} = 10^{51}$ 
   erg is the total energy released by a typical SN explosion, 
   $E_{\mathrm{wind}} = 10^{49}$ erg is the total energy injected into the ISM 
   by a typical massive star through stellar winds. $R_{\mathrm{SNII}}(t)$ and 
   $R_{\mathrm{SNIa}}(t)$ are the rates of SN explosions, which are computed 
   according to Matteucci \& Greggio (1986). $\eta_{\mathrm{SNII}}$, 
   $\eta_{\mathrm{SNIa}}$ and $\eta_{\mathrm{wind}}$ represent the 
   efficiencies with which the energies of the stellar explosions and stellar 
   winds are transferred into the ISM. According to Recchi et al.'s (2001, 
   2004) chemodynamical results for I\,Zw\,18, the efficiencies of energy 
   transfer for Type II SNe, Type Ia SNe and stellar winds are 
   $\eta_{\mathrm{SNII}}$ = 0.03, $\eta_{\mathrm{SNIa}}$ = 1.0 and 
   $\eta_{\mathrm{wind}}$ = 0.03, respectively (since SNeIa explode in a 
   hotter medium, rarefied by previous SNII explosions, their energy is more 
   efficiently thermalized). However, in case of protracted SF, it is likely 
   that SNe of both types meet with the same ISM conditions. Therefore, 
   $\eta_{\mathrm{SNII}} \simeq \eta_{\mathrm{SNIa}}$ would be a more 
   reasonable choice. We test both hypotheses in our study. The binding energy 
   of the gas is a function of the assumed dark matter (DM) content and 
   distribution. A massive ($M_{\mathrm{d}}/M_{\mathrm{lum}}$ = 10) diffuse 
   ($R_{\mathrm{eff}}/R_{\mathrm{d}}$ = 0.1) dark halo is assumed, where 
   $M_{\mathrm{d}}$ and $M_{\mathrm{lum}}$ are the dark and luminous mass, 
   respectively, and $R_{\mathrm{eff}}$ and $R_{\mathrm{d}}$ are the effective 
   radius and the radius of the DM core (see Table~\ref{tabcommonpar}). Once 
   the conditions for the onset of a galactic-scale outflow are met, metals 
   are assumed to be lost more easily than the neutral hydrogen. This is 
   expected on the basis of the geometry and mass of the system: flattened 
   galaxies of 10$^8$--10$^9$ M$_\odot$ loose preferentially metal-enriched 
   matter along the polar direction, while the unprocessed gas remains almost 
   unperturbed (De Young \& Gallagher 1990; Mac Low \& Ferrara 1999; D'Ercole 
   \& Brighenti 1999; Recchi et al. 2001; Marcolini, Brighenti \& D'Ercole 
   2004). Most of the metal-rich ejecta are pushed to large distances from the 
   galaxy (of the order of several kpc), but their final fate is uncertain: 
   they could either cool and fall back on to the galaxy plane, or be removed 
   from the galaxy. Hydrodynamical simulations of the interaction between 
   SN-powered gas outflows and the local intergalactic medium (IGM) in 
   fiducial galaxies resembling dwarf spheroidals show that ram pressure and 
   tidal stripping can efficiently pull away matter loosely bound to the 
   galaxy (Murakami \& Babul 1999). If a flattened rather than a spherical ISM 
   distribution is assumed, the fraction of metals retained by a moving galaxy 
   is found to be up to three times larger than that retained by a galaxy at 
   rest (Marcolini et al. 2004). This is interesting in the light of the 
   empirical evidence that dwarf spheroidals are depleted of their gas, 
   whereas DIGs and BCDs still show large gaseous contents.

   We follow in detail the time evolution of several chemical species (H, D, 
   He, C, N, O, Mg, Si, S, Ca and Fe), by means of the following basic 
   equation:
   \begin{equation}
   \frac{{\mathrm{d}}{\mathcal{G}}_{i}(t)}{{\mathrm{d}}t} = 
    - X_{i}(t)\psi(t) + {\mathcal{R}}_{i}(t) 
     + \frac{{\mathrm{d}}{\mathcal{G}}_{i}^{\mathrm{inf}}(t)}{{\mathrm{d}}t}
      - \frac{{\mathrm{d}}{\mathcal{G}}_{i}^{\mathrm{out}}(t)}{{\mathrm{d}}t},
       \label{eqEvol}
   \end{equation}
   where ${\mathcal{G}}_{i}(t) = X_{i}(t)M_{\mathrm{gas}}(t)/M_{\mathrm{nor}}$ 
   is the fractional gas mass in the form of the element $i$ at the time $t$, 
   normalized to a fixed total mass, $M_{\mathrm{nor}}$. The quantity 
   $X_{i}(t)$ represents the abundance by mass of the element $i$ at the time 
   $t$ and the summation over all the elements in the gas mixture is equal to 
   unity. $\psi(t)$ is the SFR, while ${\mathcal{R}}_{i}(t)$ represents the 
   fraction of mass which is restored into the ISM in the form of the element 
   $i$ by dying stars. A detailed description of this term, which contains all 
   the prescriptions about the stellar IMF, stellar nucleosynthesis and Type 
   Ia SN progenitors, can be found in Matteucci \& Greggio (1986). The 
   specific yield sets adopted in this work are discussed in 
   Sect.~\ref{secYields}. Finally, the last two terms on the right-hand side 
   of Equation~(\ref{eqEvol}) account for possible gas inflows and outflows, 
   respectively.

   In many literature models for dwarf galaxy formation, DIGs and BCDs form 
   from continuous infall of gas of primordial chemical composition, until a 
   mass from a few 10$^8$ M$_\odot$ up to $\sim$ 10$^9$ M$_\odot$ is 
   accumulated. The rate of gas infall is normally parametrized as
   \begin{equation}
   \frac{{\mathrm{d}}{\mathcal{G}}_{i}^{\mathrm{inf}}(t)}{{\mathrm{d}}t} = 
    \frac{X_{i}^{\mathrm{inf}} e^{-t/\tau}}{\tau (1 - 
     e^{-t_{\mathrm{now}}/\tau})},
       \label{eqInf}
   \end{equation}
   with $\tau$, the infall time scale, being the most critical free parameter 
   and largely varying amongst different authors (e.g., $\tau$ = 0.5 Gyr for 
   Matteucci \& Tosi's 1985 models; $\tau$ = 1, 4, 7 Gyr for Mouhcine \& 
   Contini's 2002 models). Here we run models with the choices for $\tau$ 
   specified in Table~\ref{tabcommonpar}. $X_{i}^{\mathrm{inf}} = 
   X_{i}^{\mathrm{P}}$ is the primordial abundance of element $i$ in the 
   inflowing gas. Here we adopt (D/H)$_{\mathrm{P}}$ = 2.5~$\times$~10$^{-5}$, 
   ($^3$He/H)$_{\mathrm{P}}$ = 0.9~$\times$~10$^{-5}$, and $Y_{\mathrm{P}}$ = 
   0.248 (see Romano et al. 2003, and references therein). The quantity 
   $t_{\mathrm{now}}$, the present time, is set to 13.7 Gyr. In addition, we 
   test also the cases of a constant infall rate,
   \begin{equation}
   \frac{{\mathrm{d}}{\mathcal{G}}_{i}^{\mathrm{inf}}(t)}{{\mathrm{d}}t} = 
    \mathrm{const},
       \label{eqInf}
   \end{equation}
   and of an increasing exponential,
   \begin{equation}
   \frac{{\mathrm{d}}{\mathcal{G}}_{i}^{\mathrm{inf}}(t)}{{\mathrm{d}}t} = 
    \frac{X_{i}^{\mathrm{inf}} e^{t/\tau}}{\tau (e^{t_{\mathrm{now}}/\tau} - 
     1)},
       \label{eqInf}
   \end{equation}
   with the lowest accretion rates at the early stages of galaxy formation and 
   the highest ones at later epochs, to coincide with the observed late SF 
   boosting. Notice that the infall rate is normalized to the quantity 
   $M_{\mathrm{nor}}$, which obeys the condition: $M_{\mathrm{nor}}$~= 
   $M_{\mathrm{gas}}(t_{\mathrm{now}}) + M_{\mathrm{stars}}(t_{\mathrm{now}}) 
   + M_{\mathrm{rem}}(t_{\mathrm{now}}) + M_{\mathrm{GW}}(t_{\mathrm{now}})$, 
   where $M_{\mathrm{gas}}(t_{\mathrm{now}})$, 
   $M_{\mathrm{stars}}(t_{\mathrm{now}})$, and 
   $M_{\mathrm{rem}}(t_{\mathrm{now}})$ are the masses in form of gas, stars 
   and stellar remnants, respectively, and $M_{\mathrm{GW}}(t_{\mathrm{now}})$ 
   is the total mass ejected from the galaxy by GWs at the present time.

   The rate of gas loss via GWs at the time $t$ is assumed to be proportional 
   to the amount of gas present at that time. For each element $i$,
   \begin{equation}
   \frac{{\mathrm{d}}{\mathcal{G}}_{i}^{\mathrm{out}}(t)}{{\mathrm{d}}t} = 
    w_{i}(t) X_{i}^{\mathrm{out}}(t) {\mathcal{G}}(t),
   \end{equation}
   where $X_{i}^{\mathrm{out}}(t) = X_{i}(t)$. Here we call for a dependence 
   of $w_{i}(t)$ on time via a dependence on the Type II plus Type Ia SN rate 
   (expressed in number Gyr$^{-1}$):
   \begin{equation}
   w_{i}(t) = \beta_i [\eta_{\mathrm{SNII}} R_{\mathrm{SNII}}(t) + 
              \eta_{\mathrm{SNIa}} R_{\mathrm{SNIa}}(t)].
   \label{eqWind}
   \end{equation}
   The parameter $\beta_i$ is a proportionality constant indicating the 
   efficiency of mass loss for a specific element $i$. In the case of 
   differential winds, the value of $\beta_{i}$ is assumed to be different for 
   different elements. In particular, higher $\beta_{i}$ values are assigned 
   to heavy elements produced in SNe, while only a minor fraction of the 
   neutral hydrogen is allowed to leave the galaxy.

   The SFR has the simple form:
   \begin{equation}
     \psi(t) = \nu {\mathcal{G}}(t),
   \end{equation}
   where $\nu$, the efficiency of SF, is expressed in units of Gyr$^{-1}$, and 
   is fixed so as to recover the SFR inferred from the observations.

   The adopted IMF (by mass) is an extension of that originally proposed by 
   Salpeter (1955) to the whole stellar mass range:
   \begin{equation}
     \varphi(m) = {\mathcal{C}} m^{-x},
   \end{equation}
   with $x = 1.35$ for 0.1 $\le m/$M$_\odot \le$ 100. $\mathcal{C}$ is the 
   normalization constant, and is obtained by imposing the condition
   \begin{equation}
     \int_{0.1}^{100} \varphi(m) {\mathrm{d}}m = 1.
   \end{equation}
   This IMF is in agreement with the results of the CMD analysis for both 
   NGC\,1705 and NGC\,1569, though one must be aware that \emph{the synthetic 
   CMDs actually constrain the IMF slope of these galaxies only for ${\itl m 
   \ga}$ 6 M$_\odot$} (Annibali et al. 2003). Flatter slopes in the low-mass 
   range cannot be excluded (Angeretti et al. 2005). We discuss the effect on 
   model results of adopting different IMF formulations -- still consistent 
   with the CMD analysis -- in Sect.~\ref{secresIMF}.

   \subsection{Nucleosynthesis prescriptions} \label{secYields}

%%%%%%%%%%%%%%%%%%%%%%%%%%%%%%%%%%%%%%%%%%%%%%%%
%
   \begin{table*}
   \begin{minipage}{17.5cm}
   \caption{SFH, GW efficiency and nucleosynthesis prescriptions.}
   \label{tabpar}
   \begin{tabular}{@{}lcccc@{}}
   \hline
   \multicolumn{5}{c}{Star formation history} \\
   Model & Number of bursts & Time of burst occurrence & Duration of the burst 
     & $\nu$ \\
         &                  & (Gyr)                    & (Gyr) 
     & (Gyr$^{-1}$) \\
   \cline{2-5}
%   \textsl{1} & 2 & 8.7/13.7                     & 5/ongoing
%             & 1--3/17 \\
   \textsl{1} & \multicolumn{4}{c}{Star formation history from Annibali et al. 
                                  (2003)} \\
   \textsl{2} & 3 & 5.55/11.55/13.6              & 6/2/0.1            
             & 0.05/0.15/2 \\
   \textsl{3} & 3 & 5.55/12.55/13.6              & 6/1/0.1            
             & 0.05/0.3/3 \\
   \textsl{4} & 5 & 4.575/6.575/8.575/11.55/13.6 & 0.25/0.25/0.25/2/0.1 
             & 0.5/0.5/0.5/0.15/2 \\
   \textsl{5} & 3 & 0.7/11.55/13.6                 & 10.5/2/0.1           
             & 0.012/0.15/2.4 \\
   \hline
   \multicolumn{5}{c}{GW ejection efficiencies ($\beta_i$)$^{\mathrm{a}}$} \\
         & \multicolumn{2}{c}{H, D, He} & \multicolumn{2}{c}{C, N, O, and 
         heavier} \\
   \cline{2-5}
   \textsl{a}    & \multicolumn{2}{c}{2   $\times$ 10$^{-6}$} & 
                   \multicolumn{2}{c}{2   $\times$ 10$^{-5}$} \\
   \textsl{b}    & \multicolumn{2}{c}{2   $\times$ 10$^{-6}$} & 
                   \multicolumn{2}{c}{2.5 $\times$ 10$^{-5}$} \\
   \textsl{c}    & \multicolumn{4}{c}{As above, except for nitrogen, for which 
                   $\beta_{\mathrm{N}}$ = 5.5 $\times$ 10$^{-5}$} \\
   \textsl{d}    & \multicolumn{2}{c}{0.5 $\times$ 10$^{-5}$} & 
                   \multicolumn{2}{c}{2.5 $\times$ 10$^{-5}$} \\
   \textsl{e}    & \multicolumn{2}{c}{0.5 $\times$ 10$^{-5}$} & 
                   \multicolumn{2}{c}{1.5 $\times$ 10$^{-5}$} \\
   \textsl{f}    & \multicolumn{4}{c}{As case \textsl{d}, except for nitrogen, 
                   for which $\beta_{\mathrm{N}}$ = 5.5 $\times$ 10$^{-5}$} \\
   \hline
   \multicolumn{5}{c}{Nucleosynthesis prescriptions} \\
         & \multicolumn{2}{c}{LIMS} 
         & \multicolumn{2}{c}{Massive stars} \\
   \cline{2-5}
   \textsl{N} & \multicolumn{2}{c}{van den Hoek \& Groenewegen 1997, 
                $\eta_{\mathrm{AGB}}$ const} 
             & \multicolumn{2}{c}{Nomoto et al. 1997$^{\mathrm{b}}$} \\
   \textsl{W} & \multicolumn{2}{c}{van den Hoek \& Groenewegen 1997, 
                $\eta_{\mathrm{AGB}}$ var} 
             & \multicolumn{2}{c}{Woosley \& Weaver 1995$^{\mathrm{b}}$} \\
   \textsl{H} & \multicolumn{2}{c}{van den Hoek \& Groenewegen 1997, 
                minimum HBB} 
             & \multicolumn{2}{c}{Nomoto et al. 1997$^{\mathrm{b}}$} \\
   \textsl{M} & \multicolumn{2}{c}{Meynet \& Maeder 2002$^{\mathrm{c}}$} 
             & \multicolumn{2}{c}{Meynet \& Maeder 2002$^{\mathrm{c}}$} \\
   \hline
   \end{tabular}
   \begin{list}{}{}
   \item[$^{\mathrm{a}}$] Notice that the actual GW ejection efficiencies are 
                          given by $w_{i}(t) = \beta_i [\eta_{\mathrm{SNII}} 
			  R_{\mathrm{SNII}}(t) + \eta_{\mathrm{SNIa}} 
			  R_{\mathrm{SNIa}}(t)]$, where $R_{\mathrm{SNII}}(t)$ 
			  and $R_{\mathrm{SNIa}}(t)$ are the Type II and Type 
			  Ia SN rates by number at time $t$, respectively, 
			  each one multiplied by the appropriate SN heating 
			  efficiency.
   \item[$^{\mathrm{b}}$] The yields of $^{12}$C for stars more massive than 
                          40 M$_\odot$ were multiplied by a factor of 3 in 
                          order to fit the Milky Way carbon abundance data 
                          (Chiappini et al. 2003a).
   \item[$^{\mathrm{c}}$] We adopt the set of yields computed for initial 
                          rotational velocities of the stars of 
                          $v_{\mathrm{ini}}$ = 300 km s$^{-1}$.
   \end{list}
   \end{minipage}
   \end{table*}
%
%%%%%%%%%%%%%%%%%%%%%%%%%%%%%%%%%%%%%%%%%%%%%%%%

   To quantitatively estimate the uncertainty on the results related to 
   different nucleosynthesis assumptions, we have run our models assuming 
   various sets of stellar yields. 

   For low- and intermediate-mass stars (LIMS), i.e., stars with 0.8 $\le 
   m$/M$_\odot$ $\le$ 8, both the yields of van den Hoek \& Groenewegen (1997) 
   and those computed by Meynet \& Maeder (2002) for rotating stellar models 
   are considered. For the van den Hoek \& Groenewegen (1997) yields, we adopt 
   either those computed with constant mass loss parameter along the 
   asymptotic giant branch (AGB), $\eta_{\mathrm{AGB}}$ = 4 
   (Table~\ref{tabpar}; models labelled $N$), or those computed with 
   metallicity-dependent $\eta_{\mathrm{AGB}}$ (Table~\ref{tabpar}; models 
   labelled $W$), or those computed with the minimum hot bottom burning (HBB) 
   extent allowed by the observations (Table~\ref{tabpar}; models labelled 
   $H$). For the Meynet \& Maeder (2002) yields, we adopt those computed for 
   initial rotational velocities of the stars of $v_{\mathrm{ini}}$ = 300 km 
   s$^{-1}$ (Table~\ref{tabpar}; models labelled $M$).

   For massive stars, i.e., stars with $m >$ 8 M$_\odot$, the adopted 
   nucleosynthesis prescriptions are either from Nomoto et al. (1997; 
   Table~\ref{tabpar}; models $N$ and $H$), or from Woosley \& Weaver (1995; 
   Table~\ref{tabpar}; models labelled $W$), or from Meynet \& Maeder (2002; 
   Table~\ref{tabpar}; models labelled $M$).

   Type Ia SN nucleosynthesis is accounted for by adopting the prescriptions 
   by Thielemann, Nomoto \& Hashimoto (1993), their model W7.

   Each model is characterized by one number and one pair of letters; the 
   number refers to the adopted SFH, the small letter to the adopted GW 
   efficiency (which, according to the differential wind picture, is higher 
   for the heavy elements than for the light ones), the capital letter to the 
   adopted stellar yields (see Table~\ref{tabpar}).

   \section{Results}

%%%%%%%%%%%%%%%%%%%%%%%%%%%%%%%%%%%%%%%%%%%%%%%% PLEASE, TWO COLUMNS FIGURE!!!
%
   \begin{figure*}
   \psfig{figure=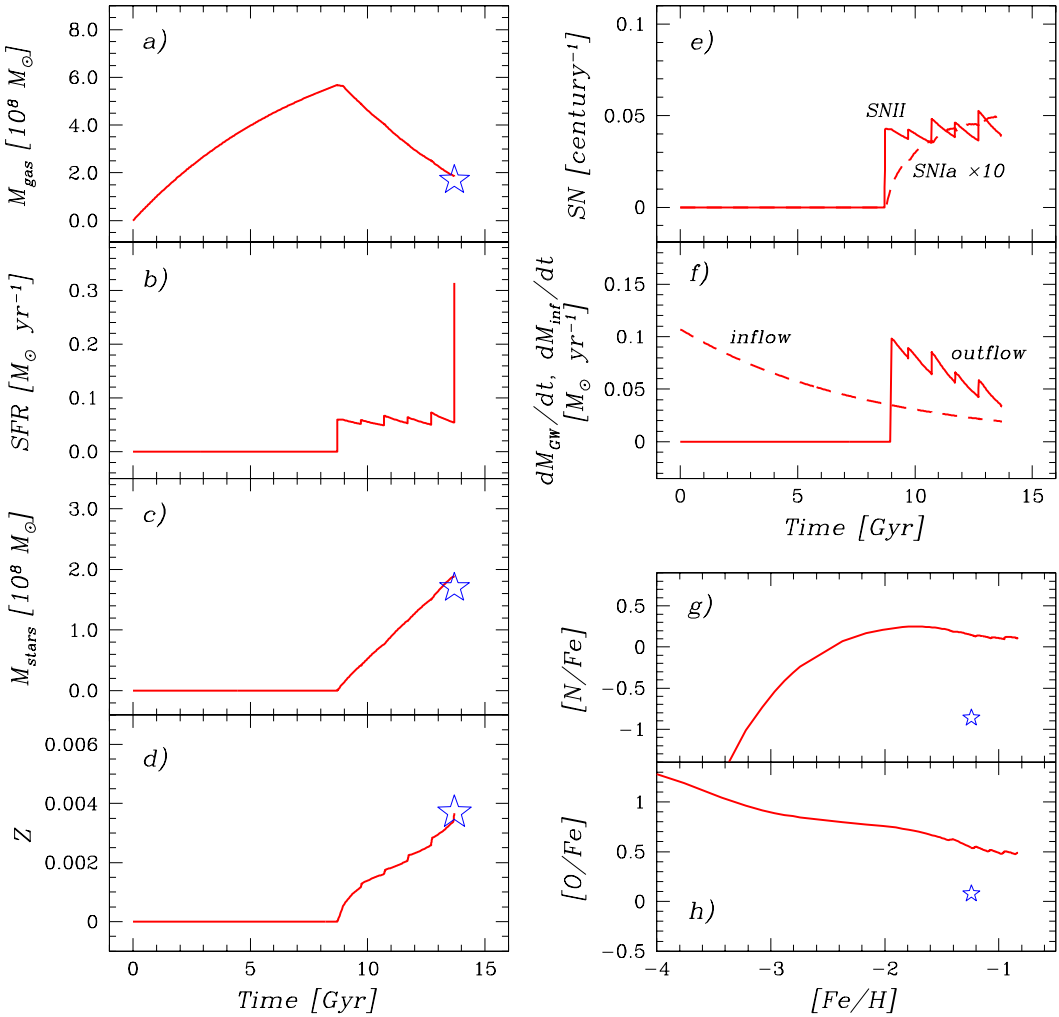,width=15cm}
      \caption{ a) $M_{\mathrm{gas}}$, b) SFR, c) $M_{\mathrm{stars}}$, d) gas 
	        metallicity, e) Type II \emph{(solid line)} and Type Ia 
		\emph{(dashed line;} multiplied by a factor of 10 to make it 
		clearly visible) SN rate, f) $\dot{M}_{\mathrm{GW}}$ 
		\emph{(solid line)} and $\dot{M}_{\mathrm{inf}}$ \emph{(dashed 
		line)} as functions of time; g) [N/Fe] and h) [O,Fe] as 
		functions of [Fe/H] for Model~\textsl{1aN} for NGC\,1705. The 
		stars represent observed values. Both the theoretical and the 
		observed (Aloisi et al. 2005) [el/Fe], [Fe/H] ratios are 
		normalized to solar values by Asplund et al. (2004). 
             }
         \label{fig17051aN}
   \end{figure*}
%
%%%%%%%%%%%%%%%%%%%%%%%%%%%%%%%%%%%%%%%%%%%%%%%%

   Most GCE studies published so far assume that DIGs and BCDs evolve through 
   successive bursts of SF -- which possibly induce GWs -- alternating long 
   quiescency periods (Matteucci \& Chiosi 1983; Matteucci \& Tosi 1985; 
   Pilyugin 1993; Marconi et al. 1994; Bradamante et al. 1998). Very few 
   attempts are made at modelling specific objects (e.g., Carigi et al. 1999; 
   Recchi et al. 2001, 2004, 2005). However, given the high number of free 
   parameters involved in dwarf galaxy modelling, it is instead of fundamental 
   importance to model individual galaxies whose SFH and IMF have been derived 
   from observations.

   \subsection{NGC\,1705} \label{secRes1705}

   NGC\,1705 has its SFH and IMF slope known thanks to deep \textsl{HST} 
   photometry probing epochs up to 5 Gyr ago (see Sect.~\ref{sec1705}). The 
   stellar mass astrated at the inferred SFR over the last 5 Gyrs is 
   sufficient to account for the stellar mass we observe in the galaxy today 
   (Annibali et al. 2003). Therefore, there is no need to speculate upon 
   earlier SF activity, and we can safely assume that the galaxy formed the 
   vast majority of its stars during the last $\sim$ 5 Gyrs at the average 
   rates suggested by Annibali et al. (2003; see Table~\ref{tab1705sfr}). 
   Following those authors, the stellar masses are distributed according to a 
   Salpeter IMF ($x$ = 1.35) over the whole (0.1--100~M$_\odot$) stellar mass 
   range, unless otherwise stated.

   In Fig.~\ref{fig17051aN} we show the adopted SFH together with some results 
   from Model~\textsl{1aN}. The adopted structural parameters of the model 
   are: $M_{\mathrm{nor}}$~= 7~$\times$~10$^8$ M$_\odot$, $R_{\mathrm{eff}}$~= 
   1.2 kpc, $M_{\mathrm{d}}$~= 7~$\times$~10$^9$~M$_\odot$, 
   $R_{\mathrm{eff}}/R_{\mathrm{d}}$~= 0.1 (see Sect.~\ref{secModels} and 
   Table~\ref{tabcommonpar} for explanations regarding model parameters). The 
   assumed infall law is exponentially decreasing with time with $e$-folding 
   time $\tau$ =~8~Gyr (Fig.~\ref{fig17051aN}f, \emph{dashed line}). This is a 
   quite long time scale, comparable to that for the formation of the solar 
   neighbourhood ($\tau_{\mathrm{SN}}$~=~7 Gyr; Chiappini, Matteucci \& 
   Gratton 1997). It allows a non-negligible fraction of the cold gas to be 
   available for SF at late times. The assumed thermalization efficiencies 
   from SNe and stellar winds are $\eta_{\mathrm{SNII}}$ = 
   $\eta_{\mathrm{SNIa}}$ = $\eta_{\mathrm{wind}}$ = 0.20.

   Model~\textsl{1aN} well reproduces the present-day gaseous 
   (Fig.~\ref{fig17051aN}a) and stellar (Fig.~\ref{fig17051aN}c) contents of 
   NGC\,1705, as well as its current metallicity (Fig.~\ref{fig17051aN}d). In 
   the framework of this model, a galactic outflow develops very soon, already 
   $\sim$ 250 million years after the onset of the SF, due to the strong SN 
   feedback. We predict a current mass loss rate of $\sim$~0.04 M$_\odot$ 
   yr$^{-1}$ through the outflow, i.e., about an order of magnitude lower than 
   the present-day SFR. This is due to the fact that SNe with progenitors born 
   during the last sudden burst of SF, have not had time to explode and leave 
   their signature on the GW properties yet. Therefore, it is still the 
   previous SF activity which determines the features of the outflow in our 
   model. An attempt to evaluate the actual mass loss rate of NGC\,1705 has 
   been made by Meurer et al. (1992) in their multi-wavelength study of this 
   object. They parameterize the rate of mass loss from the galaxy in terms of 
   two quantities, $M_{\mathrm{flow}}$, the total mass of gas entrained in the 
   outflow, and $M_{\mathrm{tot}}/L_{\mathrm{B}}$, the total mass-to-light 
   ratio. The uncertainty in $M_{\mathrm{flow}}$ (nearly three orders of 
   magnitude) and the poorly known DM contribution to $M_{\mathrm{tot}}$ make 
   the observed mass loss rate really ill-constrained: Meurer et al. indicate 
   a minimum mass loss rate of 0.0026 M$_\odot$ yr$^{-1}$ [one order of 
   magnitude below our prediction, $\dot M_{\mathrm{GW}}(t_{\mathrm{now}}) 
   \simeq$ 0.04 M$_\odot$ yr$^{-1}$], but values up to three orders of 
   magnitude higher are also permitted. It is thus clear that theoretical 
   models for the development of GWs are poorly constrained by the 
   observations currently available for NGC\,1705. In this respect, a much 
   more interesting case to study is NGC\,1569, which has its outflow much 
   better characterized by deep \textsl{Chandra} spectral imaging (Martin et 
   al. 2002). Therefore, we defer any further consideration on the GW 
   behaviour to Sect.~\ref{sec1569res}, where we treat in detail the formation 
   and evolution of NGC\,1569.

   As mentioned above, we need to assume that the winds driving the evolution 
   of NGC\,1705 are differential, i.e., they carry away a large fraction of 
   the newly produced metals, while retaining most of hydrogen. This is 
   necessary in order to reproduce both the observed present-day gas content 
   and the current metallicity of the galaxy. In Model~\textsl{1aN}, we set 
   the $\beta_i$ parameters to 2 $\times$ 10$^{-5}$ for the heavy elements and 
   to 2 $\times$ 10$^{-6}$ for the light ones (see Sect.~\ref{secModels} and 
   Table~\ref{tabpar}). Setting $\beta_i$ to 2 $\times$ 10$^{-6}$ for all the 
   chemical species would result in a present-day metallicity of 0.017, i.e. 
   almost solar. On the contrary, setting $\beta_i$ to 2 $\times$ 10$^{-5}$ 
   for all the chemical elements would result in a present-day metallicity of 
   0.04, i.e. twice solar, since in this case the metals are mixed in a much 
   lower gaseous mass: $M_{\mathrm{gas}}(t_{\mathrm{now}})~\sim 
   3~\times~10^6$~M$_\odot$, at variance with the observations. Clearly, the 
   model is not unique, and the formulation for the $w_i(t)$ parameters is 
   allowed to change. However, in agreement with previous analysis (Pilyugin 
   1993; Marconi et al. 1994; Recchi et al. 2001, 2004, 2005) we find that 
   models assuming a common value for $w_i(t)$ (i.e., no differential winds) 
   unavoidably fail to reproduce (at least some of) the observational 
   constraints.

   \subsubsection{Detailed chemical properties of the neutral and ionized gas 
                 phases in NGC\,1705} \label{sec1705chem}

\begin{figure}
   \psfig{figure=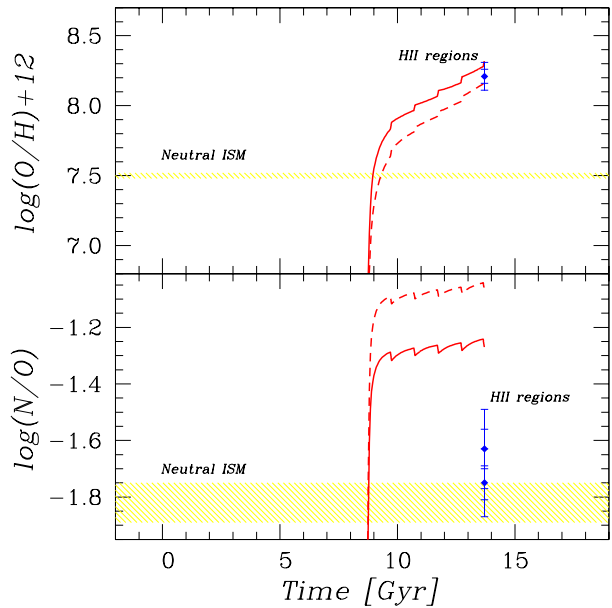,width=\columnwidth}
      \caption{ Temporal evolution of oxygen \emph{(top panel)} and nitrogen 
               to oxygen \emph{(bottom panel)} in the ISM of NGC\,1705. 
               Predictions from Model~\textsl{1aN} (\emph{solid lines:} 
	       Salpeter IMF; \emph{dashed lines:} steeper IMF -- see text for 
	       details) are compared to the available observations. The 
	       present-day oxygen abundance of NGC\,1705 is that given by Lee 
	       \& Skillman (2004; mean value of 
	       [O\,{\tiny III}]\,$\lambda$\,4363 measurements; \emph{small 
               filled diamond}). The oxygen abundance of the neutral ISM is 
               that derived from far-UV observations (Aloisi et al. 2005; 
	       \emph{dashed horizontal band}), not corrected for contamination 
	       by ionized gas lying along the line of sight. The current N/O 
	       ratio in NGC\,1705 is that suggested by Lee \& Skillman (2004; 
	       \emph{small filled diamonds}), with (lower value) and without 
	       (higher value) region B4. A value of log(N/O) = $-$1.82 $\pm$ 
	       0.07 characterizes the neutral phase (Aloisi et al. 2005; 
	       \emph{dashed horizontal band}), which is the same as H\,{\small 
	       II} regions within the errors.
             }
         \label{fig1705no}
   \end{figure}

   The abundances of oxygen and nitrogen in the local ISM of NGC\,1705 have 
   been derived from spectroscopy of H\,{\small II} regions by several groups 
   (see Sect.~\ref{sec1705}). In the following, we adopt for comparison with 
   our model predictions the up-to-date values of log(O/H)+12 and log(N/O) 
   suggested by Lee \& Skillman (2004 -- Fig.~\ref{fig1705no}, \emph{filled 
   diamonds,} with 1- and 2-$\sigma$ error bars). While the present-day oxygen 
   abundance is very well reproduced by the model (Fig.~\ref{fig1705no}, 
   \emph{upper panel,} cfr. the final points of the tracks), the current N/O 
   ratio is largely overestimated (Fig.~\ref{fig1705no}, \emph{lower panel}). 
   In Fig.~\ref{fig1705no} we show results obtained by adopting either $x$~= 
   1.35 over the whole stellar mass range \emph{(solid lines)} or $x$~= 1.6 
   above 2~M$_\odot$ (\emph{dashed lines;} Annibali et al. 2003). From now on, 
   we will always refer to models for NGC\,1705 as models computed with $x$~= 
   1.35 over the whole stellar mass range.

   Recently, \textsl{FUSE} and Space Telescope Imaging Spectrograph (STIS) 
   Echelle spectra of NGC\,1705 have allowed the determination of the metal 
   (O, Ar, Si, Mg, Al, N, and Fe) abundances also in the neutral medium of 
   NGC\,1705 (Aloisi et al. 2005). However, when comparing them to model 
   predictions, one must be aware that saturation (especially for O\,{\small 
   I}), photoionization, contamination by gas laying along the line of sight, 
   depletion into dust grains (e.g., Fe could be more easily locked into 
   grains compared to O) all could affect the derivation of the abundances 
   from the neutral gas (Aloisi et al. 2005). In Fig.~\ref{fig17051aN}g,h we 
   display [N/Fe] and [O/Fe] as functions of [Fe/H]. All the ratios are 
   normalized to the solar values of Asplund, Grevesse \& Sauval (2004). At 
   [Fe/H] $\sim-$1.2, the [N/Fe] predicted by Model~\textsl{1aN} is higher 
   than the one measured in the neutral gas. This is not unexpected on the 
   basis of the results for the ionized gas phase shown in 
   Fig.~\ref{fig1705no}. The [O/Fe] ratio is overestimated by the model as 
   well. It is worth emphasizing that, while the observed [N/H] and [Fe/H] 
   were corrected for the presence of ionized gas lying along the line of 
   sight, [O/H] was not (Aloisi et al. 2005).

%%%%%%%%%%%%%%%%%%%%%%%%%%%%%%%%%%%%%%%%%%%%%%%%
%
%
%%%%%%%%%%%%%%%%%%%%%%%%%%%%%%%%%%%%%%%%%%%%%%%%

%%%%%%%%%%%%%%%%%%%%%%%%%%%%%%%%%%%%%%%%%%%%%%%% PLEASE, TWO COLUMNS FIGURE!!!
%
   \begin{figure*}
   \psfig{figure=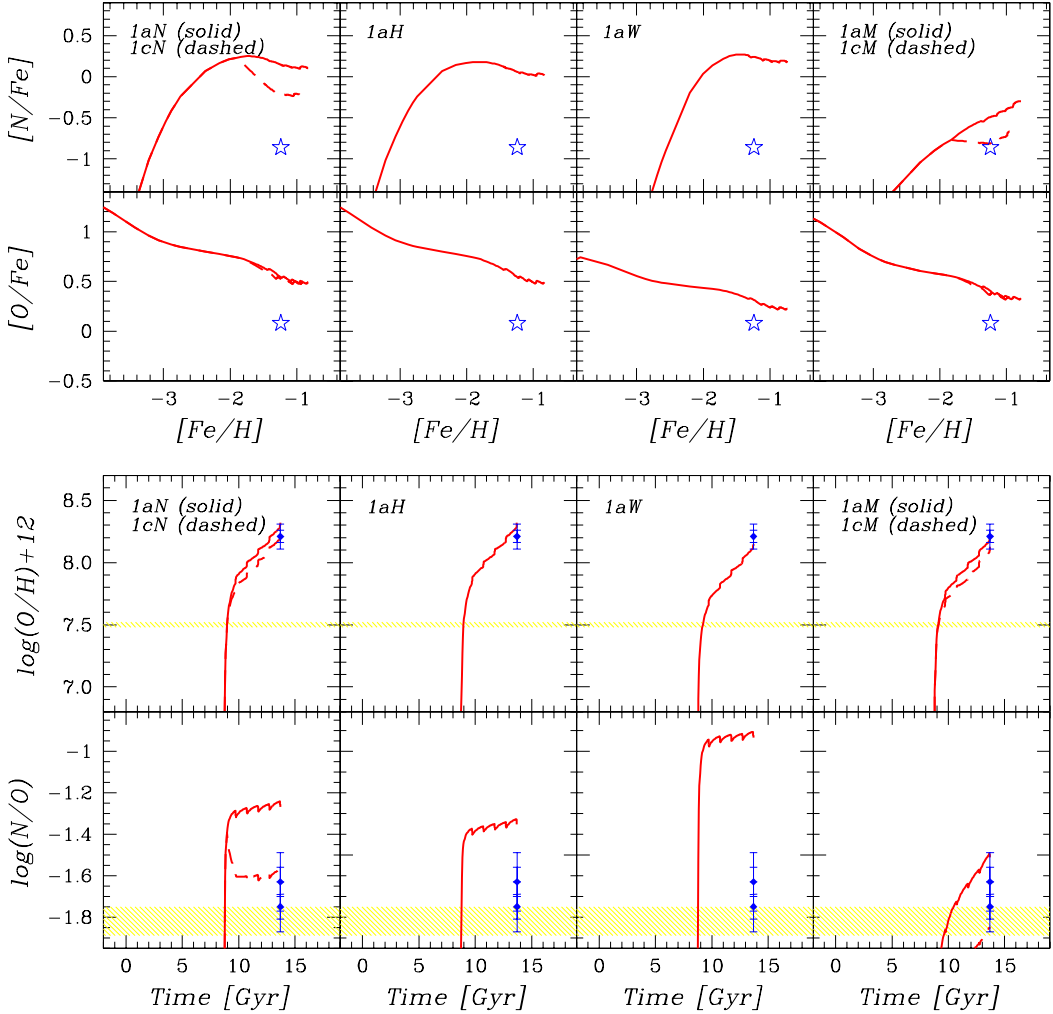,width=15cm}
      \caption{ [N/Fe] and [O/Fe] as a function of [Fe/H], log(O/H)+12 and 
	        log(N/O) as a function of time. Predictions from 
		Models~\textsl{1aN, 1aH, 1aW} and \textsl{1aM} \emph{(solid 
                lines)} as well as Models~\textsl{1cN} and \textsl{1cM} 
		\emph{(dashed lines)} are displayed for comparison one to 
		another and with data for both the neutral (Aloisi et al. 
		2005; \emph{stars} and \emph{dashed strips}) and ionized gas 
		(Lee \& Skillman 2004; \emph{small filled diamonds} with 1- 
		and 2-$\sigma$ error bars).
             }
         \label{fig1705yields}
   \end{figure*}
%
%%%%%%%%%%%%%%%%%%%%%%%%%%%%%%%%%%%%%%%%%%%%%%%%

   The pronounced discrepancy between model predictions and observations for 
   nitrogen -- the predicted nitrogen abundance is sistematically higher than 
   observed, both for the neutral and the ionized gas phases -- requires 
   further analysis, especially since the model well reproduces the other 
   reliable observational constraints. In the remaining part of this section, 
   we discuss how current uncertainties in the theories of stellar evolution 
   and nucleosynthesis may affect our results.

   \paragraph{Less hot bottom burning in intermediate-mass stars?}

   In Fig.~\ref{fig1705yields} we display the evolution of [N/Fe], [O/Fe] vs. 
   [Fe/H] and log(O/H)+12, log(N/O) vs. time predicted by models for NGC\,1705 
   adopting one of the following combinations: (i) the yields from massive 
   stars by Nomoto et al. (1997) plus the yields from LIMS by van den Hoek \& 
   Groenewegen (1997), either with a mass loss parameter along the AGB 
   independent of metallicity -- Models~\textsl{1aN} \emph{(first column, 
   solid line)} and \textsl{1cN} \emph{(first column, dashed line)} -- or with 
   the minimum HBB extent (Model~\textsl{1aH}, \emph{second column}); (ii) the 
   yields from massive stars by Woosley \& Weaver (1995) plus the yields from 
   LIMS by van den Hoek \& Groenewegen (1997) computed with 
   metallicity-dependent mass loss parameter along the AGB 
   (Model~\textsl{1aW}, \emph{third column}); (iii) stellar yields by Meynet 
   \& Maeder (2002) for rotating stellar models (Models~\textsl{1aM} and 
   \textsl{1cM}, \emph{fourth column, solid} and \emph{dashed lines,} 
   respectively). We recall that Meynet \& Maeder (2002) have not computed the 
   third dredge-up and HBB phases; hence their stellar models for LIMS are 
   somewhat incomplete. Model~\textsl{1cN} \textsl{(1cM)} differs from 
   Model~\textsl{1aN} \textsl{(1aM)} only in the adopted values of the GW 
   efficiencies. Models~\textsl{1cN} and \textsl{1cM} are discussed in detail 
   in Sect.~\ref{secInfall}.

%%%%%%%%%%%%%%%%%%%%%%%%%%%%%%%%%%%%%%%%%%%%%%%% PLEASE, TWO COLUMNS FIGURE!!!
%
   \begin{figure*}
   \psfig{figure=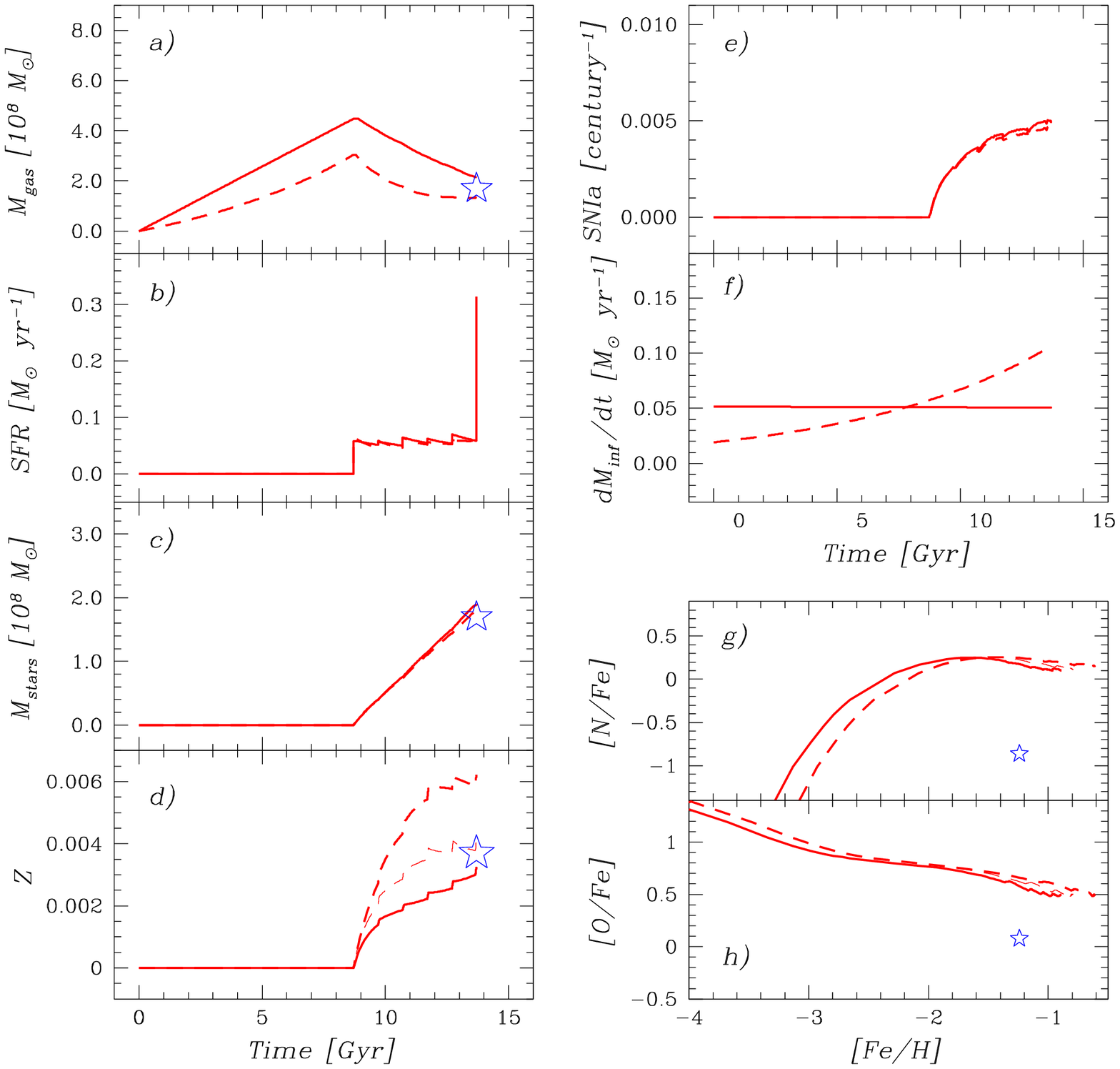,width=15cm}
      \caption{ Same as Fig.~\ref{fig17051aN}, but for models for NGC\,1705 
	        where the infall rate is either constant \emph{(thick solid 
		lines)} or exponentially increasing in time \emph{(thick} and 
		\emph{thin dashed lines)}. In the latter case, the hypothesis 
		of differential GWs may be relaxed (see text for details).
             }
         \label{figInf}
   \end{figure*}
%
%%%%%%%%%%%%%%%%%%%%%%%%%%%%%%%%%%%%%%%%%%%%%%%%

   Uncertainties in the physics of stellar models for intermediate-mass stars 
   cause the predicted log(N/O) at the present time to vary widely, from 
   $-$0.9 (Model~\textsl{1aW}) to $-$1.5 (Model~\textsl{1aM}). Stellar models 
   with rotation and without HBB give rise to the lowest theoretical N/O 
   ratios, but they are still only marginally compatible with the 
   observations. Results on the oxygen behaviour are firmer. Yet, lower [O/Fe] 
   ratios are predicted during the whole galaxy evolution when adopting the 
   yields of Woosley \& Weaver (1995) rather than those by Nomoto et al. 
   (1997) for massive stars. Meynet \& Maeder's yields are the only ones able  
   to reproduce both the observed O/H and N/O at the present time. However, 
   the discrepancies among model predictions and observations for the neutral 
   phase, though sensibly reduced, are not eliminated 
   (Fig.~\ref{fig1705yields}, \emph{last column, solid lines}).

   \paragraph{Localized massive star pollution in NGC\,1705?} \label{parSelf}

   Notwithstanding the large uncertainties affecting the nitrogen yields, the 
   discrepancy between model predictions about nitrogen evolution in NGC\,1705 
   and the relevant observations is hardly explained only in terms of the 
   adopted stellar yields. A second possibility is localized self-pollution 
   from dying young massive stars born during the last 3 Myr of SF activity. 
   According to Meynet \& Maeder (2002), the N/O ratio in the ejecta of a 
   60~M$_\odot$ star, having a lifetime of $\sim$ 3.5 million years, is about 
   log(N/O)~$\simeq -$2.2 for $Z_{\mathrm{ini}}$~= 0.004. It would be more 
   interesting to compare to the observations the log(N/O) values predicted 
   for rotating massive stars with lifetimes shorter than 3 million years, 
   but, unfortunately, no star models with initial mass larger than 
   60~M$_\odot$ are provided yet. Anyway, we notice that the N/O ratio in the 
   ejecta of massive stars with initial metallicity $Z_{\mathrm{ini}}$~= 0.004 
   and initial rotational velocity $v_{\mathrm{ini}}$~= 300~km~s$^{-1}$ is a 
   decreasing function of mass (Meynet \& Maeder 2002). Therefore, \emph{lower 
   N/O ratios are expected in case of localized massive star self-pollution 
   with respect to the general predictions of homogeneous GCE models,} which 
   may solve the discrepancy between model predictions and observations. 
   However, such a solution would not account for the low nitrogen abundance 
   measured with \textsl{FUSE}. Moreover, the constancy of log(O/H)+12 among 
   different H\,{\small II} regions issues a serious challenge to this 
   interpretation.

   \subsubsection{The r\^ole of gas flows} \label{secInfall}

   IMF variations in the range allowed by CMD analyses do not allow to solve 
   the `nitrogen problem' in NGC\,1705 (see Fig.~\ref{fig1705no}, \emph{lower 
   panel}). We are thus left with the possibility that the efficiency of 
   nitrogen ejection through the wind of NGC\,1705 is higher than that for 
   oxygen, i.e., $\beta_{\mathrm N}/\beta_{\mathrm O} \ne 1$. Though this 
   hypothesis needs to be tested by detailed hydrodynamical computations 
   before it can be accepted or rejected, we compute two models for NGC\,1705 
   in which a higher $\beta_{\mathrm N}/\beta_{\mathrm O}$ value is assumed, 
   namely, $\beta_{\mathrm N}/\beta_{\mathrm O} \simeq$ 2 rather than 1 
   (Models~\textsl{1cN} and \textsl{1cM}), and find that the latter matches 
   fairly well the available observations (Fig.~\ref{fig1705yields}, 
   \emph{last column, dashed lines}). In conclusion, it seems that the 
   outflowing gas should be differentially enriched by the products of stellar 
   nucleosynthesis in order to make the model predictions match the 
   observations.

   At this point, one must check if our conclusions change by modifying the 
   prescriptions on gas inflows. Up to now, we assumed that some unprocessed 
   gas is continuosly accreted by the galaxy during the whole Hubble time 
   according to an infall law exponentially decreasing with time\footnote{This 
   is a widely adopted, but completely arbitrary, assumption of GCE models 
   (see, e.g., Mouhcine \& Contini 2002 and discussion therein).} and with a 
   time scale $\tau$~= 8~Gyr. In the following, we discuss the results 
   obtained when considering an infall rate either constant or exponentially 
   increasing with time (again with $\tau$~= 8~Gyr). The latter case would 
   simulate important mergers with gaseous lumps at late times.

%%%%%%%%%%%%%%%%%%%%%%%%%%%%%%%%%%%%%%%%%%%%%%%% PLEASE, TWO COLUMNS FIGURE!!!
%
   \begin{figure*}
   \psfig{figure=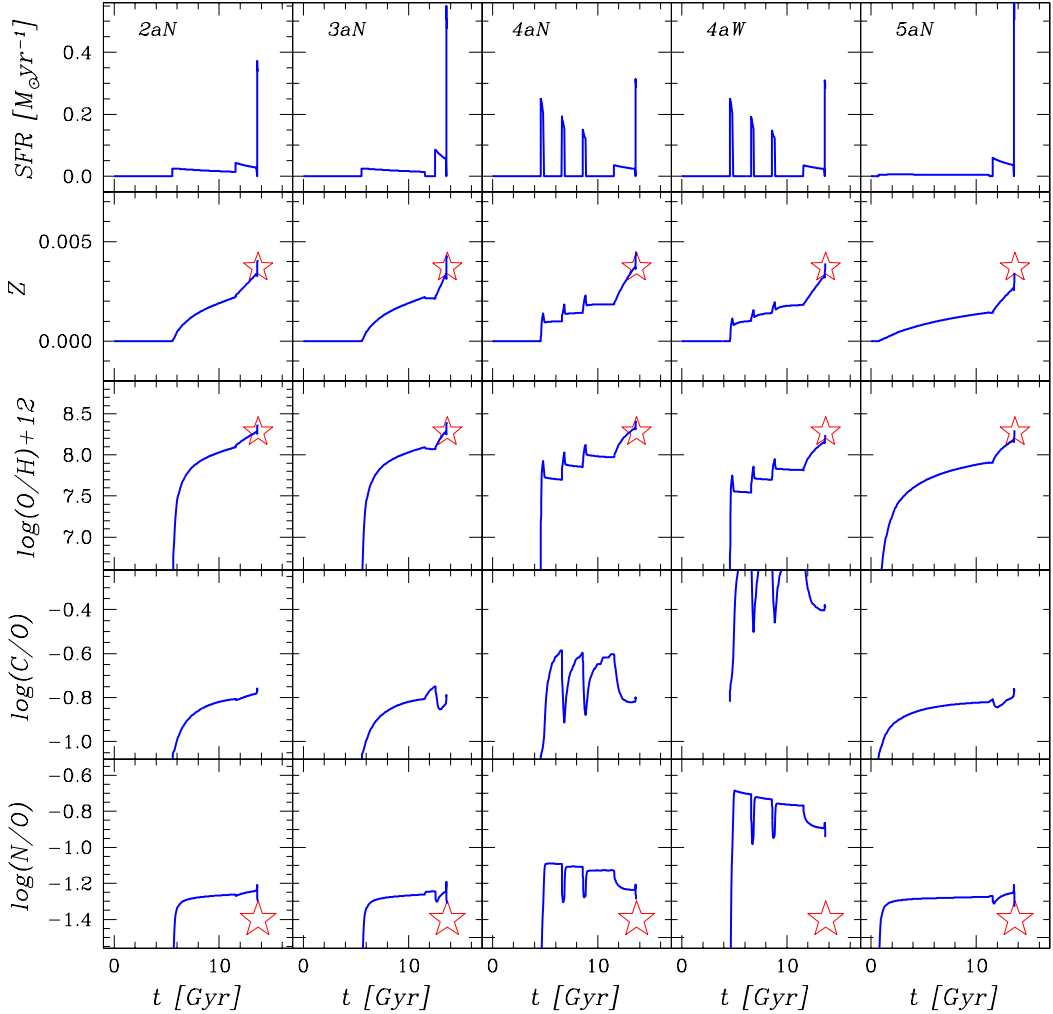,width=15cm}
      \caption{ SFR, global metallicity and CNO abundances as functions of 
                time according to Models~\textsl{2aN}, \textsl{3aN}, 
		\textsl{4aN}, \textsl{4aW} and \textsl{5aN} for NGC\,1569. See 
		text for details.
             }
         \label{fig1569res}
   \end{figure*}
%
%%%%%%%%%%%%%%%%%%%%%%%%%%%%%%%%%%%%%%%%%%%%%%%%

%%%%%%%%%%%%%%%%%%%%%%%%%%%%%%%%%%%%%%%%%%%%%%%%
%
   \begin{figure}
   \psfig{figure=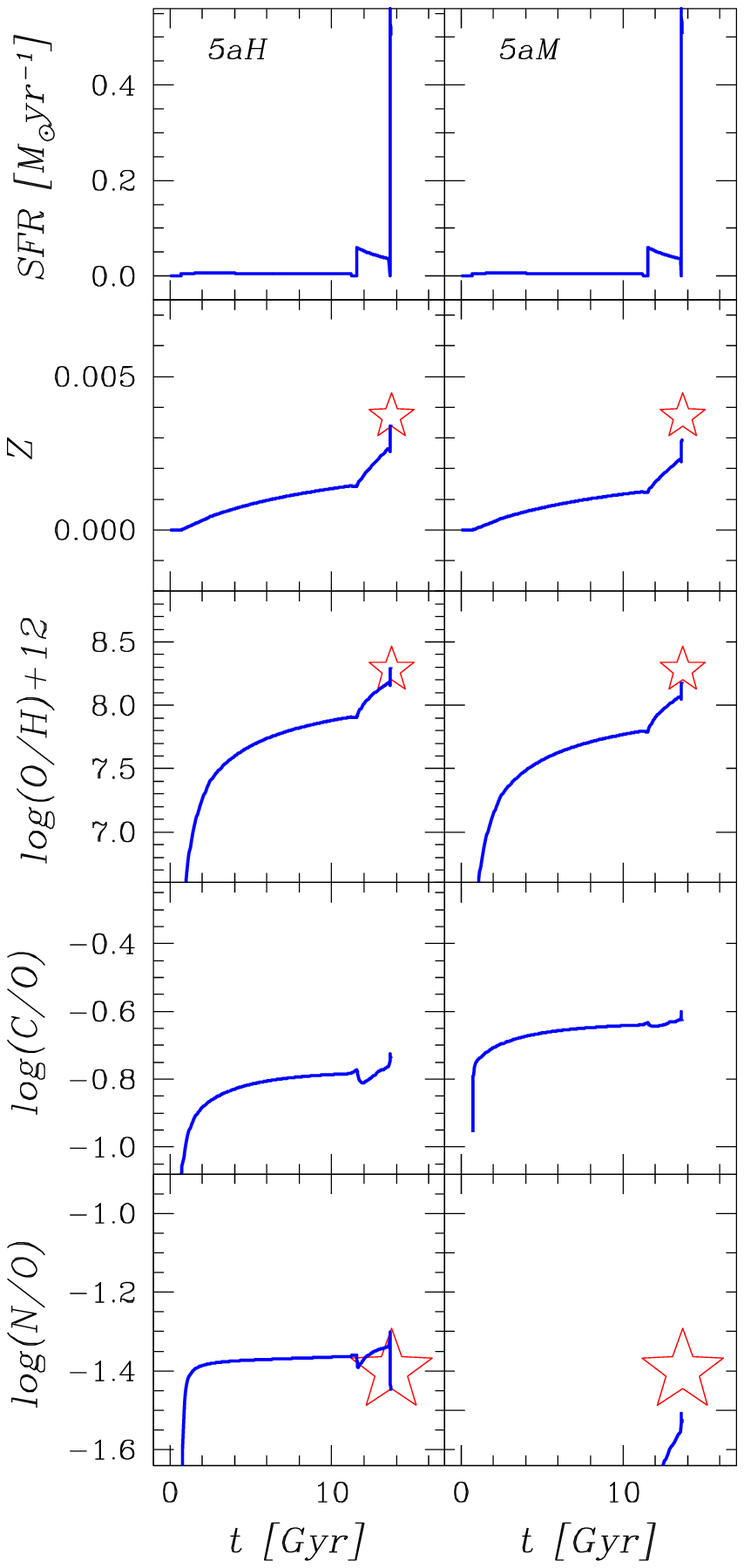,width=7cm}
      \caption{ Same as Fig.~\ref{fig1569res}, but for Models~\textsl{5aH}, 
	        \textsl{5aM}.
             }
         \label{fig1569moreres}
   \end{figure}
%
%%%%%%%%%%%%%%%%%%%%%%%%%%%%%%%%%%%%%%%%%%%%%%%%

   Results of models in which either a constant or an increasing infall rate 
   are considered are sketched in Fig.~\ref{figInf}. Once a constant infall 
   rate is assumed, predictions from Model~\textsl{1aN} (Fig.~\ref{figInf}, 
   \emph{solid lines}) almost do not differ from those obtained by the same 
   model with a mass accretion rate exponentially decreasing in time (cfr. 
   Fig.~\ref{fig17051aN}), except for the current mass loss rate, $\dot 
   M_{\mathrm{GW}}(t_{\mathrm{now}})$~= 0.1~M$_\odot$~yr$^{-1}$, rather than 
   0.04 M$_\odot$ yr$^{-1}$. In the case of constant infall, metals are mixed 
   in a slightly higher gaseous mass [$M_{\mathrm{gas}}(t_{\mathrm{now}})$~= 
   2.1~$\times$ 10$^8$ M$_\odot$ rather than 1.9~$\times$~10$^8$~M$_\odot$], 
   and one gets $Z$~= 0.0032 rather than $Z$~= 0.0037 at the present time, a 
   negligible difference. Instead, if the rate of mass accretion increases in 
   time (see Fig.~\ref{figInf}f, \emph{dashed line}), the current gaseous mass 
   of the galaxy is overestimated by the model, unless one does not raise the 
   efficiency of hydrogen loss through the wind. In Fig.~\ref{figInf}a we 
   display the behaviour of the gas mass with time predicted by 
   Models~\textsl{1dN} and \textsl{1eN}, both assuming $\beta_i$~= 
   0.5~$\times$~10$^{-5}$ for the light elements \emph{(dashed lines)}. This 
   is the highest $\beta_i$ value for hydrogen ejection we can impose while 
   still retaining enough mass to match the observations. The current 
   metallicity of the ISM is a further constraint to be fulfilled, and it 
   restricts the possible choices of $\beta_i$ for the heavy elements. As an 
   example, in Fig.~\ref{figInf}d we show results obtained with $\beta_i$ 
   equal to either 2.5 $\times$ 10$^{-5}$ (Model~\textsl{1dN}, \emph{thin 
   dashed line}) or 1.5 $\times$ 10$^{-5}$ (Model~\textsl{1eN}, \emph{thick 
   dashed line}) for elements heavier than helium. It is immediately seen that 
   an infall rate exponentially increasing in time allows one to relaxe the 
   differential wind hypothesis: 
   $\beta_{\mathrm{heavy}}/\beta_{\mathrm{light}}$~= 10 for 
   Model~\textsl{1aN}; $\beta_{\mathrm{heavy}}/\beta_{\mathrm{light}}$~= 5 for 
   Model~\textsl{1dN} (see also Table~\ref{tabpar}). At this point, one might 
   wonder whether a fine tuning of model parameters would make the 
   $\beta_{\mathrm{heavy}}/\beta_{\mathrm{light}}$ value further approach 
   unity, eventually getting rid of the need for metal-enriched GWs. We ran 
   several models, but we never reproduced simultaneously both the current 
   gaseous mass and ISM metallicity with 
   $\beta_{\mathrm{heavy}}/\beta_{\mathrm{light}}$~= 1, unless making some 
   very \emph{ad hoc} assumptions with possible huge episodes of mass 
   accretion at late times. However, NGC\,1705's outer optical isophotes 
   appear round and regular, suggesting a relatively peaceful environment 
   (Meurer et al. 1998). Finally, notice that changing the infall law does not 
   solve the above-mentioned discrepancy between model predictions and 
   observations for nitrogen: models with 
   $\beta_{\mathrm{N}}/\beta_{\mathrm{O}}$ = 1 never match the data, 
   independently of the adopted infall rate formulation (cfr. 
   Fig.~\ref{figInf}g,h and Fig.~\ref{fig17051aN}g,h). The current rate of 
   mass loss predicted by models with rate of gas accretion exponentially 
   increasing with time is $\dot M_{\mathrm{GW}}(t_{\mathrm{now}})$~= 
   0.2~M$_\odot$~yr$^{-1}$, owing to the higher hydrogen mass entrained in the 
   outflow in this case.

   \subsection{NGC\,1569} \label{sec1569res}

%%%%%%%%%%%%%%%%%%%%%%%%%%%%%%%%%%%%%%%%%%%%%%%%
%
   \begin{table*}
   \caption{Model results for NGC\,1569 (at $t = t_{\mathrm{now}} =$ 13.7 
            Gyr; \emph{second} to \emph{twelfth cols.}). The time of the onset 
	    of the first GW episode, $t_{\mathrm{GW}}$, is also shown 
	    \emph{(last col.)}.}
   \label{tab1569res}
   \begin{tabular}{@{}p{.5cm}cccccccccccc@{}}
   \hline
   Model & $M_{\mathrm{gas}}$ & $M_{\mathrm{stars}}$ & $M_{\mathrm{tot}}$ 
        & $\mu^{\mathrm{a}}$ & SFR & $Z$ & (He/H) & log(O/H)+12 & log(N/O) & 
       log(C/O) & $\dot{M}_{\mathrm{GW}}$ & $t_{\mathrm{GW}}$ \\
         & (M$_\odot$)        & (M$_\odot$)          & (M$_\odot$) &
           & (M$_\odot$ yr$^{-1}$) &     &        &             &          & 
                & (M$_\odot$ yr$^{-1}$)   & (Gyr) \\
   \hline
   \textsl{2aN} & 1.7 $\times$ 10$^8$ & 1.4 $\times$ 10$^8$ & 3.1 $\times$ 
     10$^8$ & .55 & 0.3 & 0.0041 & 0.086 & 8.37 & $-$1.30 & $-$0.77 & 0.2 & 
     6.00 \\
   \textsl{3aN} & 1.6 $\times$ 10$^8$ & 1.4 $\times$ 10$^8$ & 3.1 $\times$ 
     10$^8$ & .53 & 0.5 & 0.0043 & 0.086 & 8.40 & $-$1.31 & $-$0.80 & 0.3 & 
     6.00 \\
   \textsl{4aN} & 1.4 $\times$ 10$^8$ & 1.3 $\times$ 10$^8$ & 2.9 $\times$ 
     10$^8$ & .53 & 0.3 & 0.0045 & 0.086 & 8.41 & $-$1.29 & $-$0.81 & 0.1 & 
     4.65 \\
   \textsl{4aW} & 1.4 $\times$ 10$^8$ & 1.3 $\times$ 10$^8$ & 3.0 $\times$ 
     10$^8$ & .53 & 0.3 & 0.0039 & 0.088 & 8.24 & $-$0.94 & $-$0.39 & 0.1 & 
     4.65 \\
   \textsl{5aN} & 2.1 $\times$ 10$^8$ & 1.2 $\times$ 10$^8$ & 3.3 $\times$ 
     10$^8$ & .63 & 0.5 & 0.0034 & 0.085 & 8.30 & $-$1.33 & $-$0.77 & 0.3 & 
     2.50 \\
   \textsl{5aH} & 2.1 $\times$ 10$^8$ & 1.2 $\times$ 10$^8$ & 3.3 $\times$ 
     10$^8$ & .63 & 0.5 & 0.0034 & 0.085 & 8.30 & $-$1.45 & $-$0.74 & 0.3 & 
     2.50 \\
   \textsl{5aM} & 2.1 $\times$ 10$^8$ & 1.2 $\times$ 10$^8$ & 3.3 $\times$ 
     10$^8$ & .63 & 0.5 & 0.0029 & 0.090 & 8.19 & $-$1.52 & $-$0.63 & 0.3 & 
     2.50 \\
   \hline
   \end{tabular}
   \begin{list}{}{}
   \item[$^{\mathrm{a}}$] $\mu$ = $M_{\mathrm{gas}}(t_{\mathrm{now}})/
                                  [M_{\mathrm{gas}}(t_{\mathrm{now}}) + 
				  M_{\mathrm{stars}}(t_{\mathrm{now}})]$
   \end{list}
   \end{table*}
%
%%%%%%%%%%%%%%%%%%%%%%%%%%%%%%%%%%%%%%%%%%%%%%%%

   NGC\,1569 is another well-studied, relatively nearby late-type dwarf, whose 
   spectacular GW has been powered by an unusually high-level SF activity (see 
   Sect.~\ref{sec1569}). Given the similarities between NGC\,1705 and 
   NGC\,1569, we find it natural to start a detailed study of NGC\,1569 and 
   see whether models adopting the formalism developed for NGC\,1705 satisfy 
   the observational constraints available for NGC\,1569 as well.
 
   Our models for NGC\,1569 all assume that two major bursts of SF have 
   occurred in the last $\sim$ 1--2 Gyr, in agreement with the inferences from 
   the observed CMDs (Greggio et al. 1998; Angeretti et al. 2005). Because of 
   observational uncertainties and since the observed CMDs are relevant to 
   small inner galactic regions -- while we need a global SFR referring to the 
   whole galactic area -- we still have some degrees of freedom when choosing 
   the actual SFRs at late times. Moreover, we must allow for some SF activity 
   at earlier times, in order to build up a total stellar mass which matches 
   that inferred from the observations. Possible choices are not completely 
   arbitrary, since they must obey two more basic, stringent constraints:
   \begin{enumerate}
   \item Most of the latest SF activity is concentrated within the area 
         surveyed by Greggio et al. (1998). Therefore, it is safe to adopt the 
	 SFR suggested by those authors, $\sim$ 0.5 M$_\odot$ yr$^{-1}$, as 
	 representative of the overall SF activity across the whole body of 
	 the galaxy. Besides this value, we also test the slightly lower one 
	 derived by Hunter \& Elmegreen (2004) from H$\alpha$ imaging of 
	 NGC\,1569, $\sim$~0.3~M$_\odot$~yr$^{-1}$.
   \item SF activity prior to that surveyed by \textsl{HST} likely proceeded 
         at much lower rates (see the discussion in Angeretti et al. 2005). 
	 Therefore, for ages between $\sim$~1--2 and $\sim$~13 Gyr ago, we 
	 assume a continuous SF activity, which proceeds at rates much smaller 
	 than assumed for the recent episodes. However, only for the sake of 
	 comparison with old `bursting' models, we also examine the case of a 
	 bursting SF regime at early times, with three bursts alternating long 
	 quiescency periods.
   \end{enumerate}
   Table~\ref{tabpar} summarizes the SFHs assumed in this paper for NGC\,1569 
   in the light of the above considerations (Models from \textsl{2} to 
   \textsl{5}). Model results are discussed below.

   \subsubsection{Bursting, gasping or continuous SF regimes?}

   \begin{table*}
   \caption{Model results for NGC\,1569 (at $t = t_{\mathrm{now}} =$ 13.7 
            Gyr; \emph{fourth} to \emph{twelfth cols.}). Model prescriptions 
            are the same of Model~\textsl{5aN} \emph{(first row)}, except for 
            the adopted SNII and SNIa thermalization efficiencies 
	    \emph{(first} and \emph{second cols.)} and, eventually, DM content 
	    (\emph{third col.;} see text for more details). The time of the 
            onset of the GW, $t_{\mathrm{GW}}$, is also indicated \emph{(last 
	    col.)}.}
   \label{tab1569eta}
   \begin{tabular}{@{}p{.4cm}p{.4cm}ccccccccccc@{}}
   \hline
   $\eta_{\mathrm{SNII}}$ & $\eta_{\mathrm{SNIa}}$ & 
   $\frac{M_{\mathrm{d}}}{M_{\mathrm{lum}}}$ & $M_{\mathrm{gas}}$ & 
   $M_{\mathrm{stars}}$ & SFR                   & $Z$ & (He/H) & log(O/H)+12 & 
   log(N/O) & log(C/O) & $\dot{M}_{\mathrm{GW}}$ & $t_{\mathrm{GW}}$ \\
                          &                        & 
                                     & (M$_\odot$)        & 
   (M$_\odot$)          & (M$_\odot$ yr$^{-1}$) &     &        &             &
            &          & (M$_\odot$ yr$^{-1}$)   & (Gyr) \\
   \hline
   0.20 & 0.20 & 10 & 2.1 $\times$ 10$^8$ & 1.2 $\times$ 10$^8$ & 0.5 & 
   0.0034 & 0.085 & 8.30 & $-$1.33 & $-$0.77 & 0.3 & 2.50 \\
   0.03 & 1.00 & 10 & 2.9 $\times$ 10$^8$ & 1.2 $\times$ 10$^8$ & 0.5 & 
   0.0048 & 0.085 & 8.44 & $-$1.29 & $-$0.79 & 0.09 & 4.15 \\
   0.50 & 0.50 & 10 & 0.8 $\times$ 10$^8$ & 1.2 $\times$ 10$^8$ & 0.5 & 
   0.0035 & 0.088 & 8.31 & $-$1.36 & $-$0.71 & 0.4 & 1.30 \\
   0.50 & 0.50 & 50 & 0.9 $\times$ 10$^8$ & 1.2 $\times$ 10$^8$ & 0.5 & 
   0.0034 & 0.088 & 8.29 & $-$1.36 & $-$0.71 & 0.4 & 2.00 \\
   0.20 & 0.20 &  5 & 2.1 $\times$ 10$^8$ & 1.2 $\times$ 10$^8$ & 0.5 & 
   0.0034 & 0.085 & 8.30 & $-$1.33 & $-$0.77 & 0.3 & 2.25 \\
   \hline
   \end{tabular}
   \end{table*}
%
%%%%%%%%%%%%%%%%%%%%%%%%%%%%%%%%%%%%%%%%%%%%%%%%

%%%%%%%%%%%%%%%%%%%%%%%%%%%%%%%%%%%%%%%%%%%%%%%%
%
   \begin{figure}
   \psfig{figure=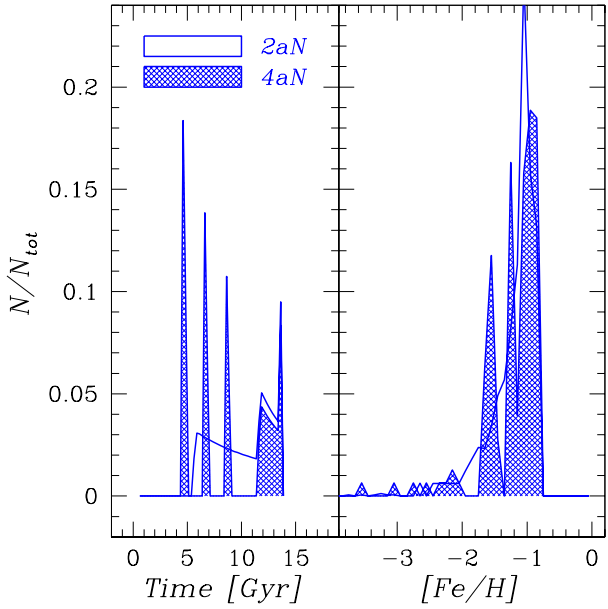,width=\columnwidth}
      \caption{ Distributions of low-mass, long-lived stars as functions of 
               time \emph{(left panel)} and metallicity \emph{(right panel)} 
               according to Models~\textsl{2aN} and \textsl{4aN} for 
	       NGC\,1569. The models assume the same SFH in the last $\sim$ 2 
	       Gyr, where more stringent constraints are available from 
	       \textsl{HST} photometry. As far as earlier activity is 
	       concerned, Model~\textsl{2aN} adopts a continuous, low-level SF 
	       over $\sim$ 6 Gyr, while Model~\textsl{4aN} assumes three 
	       short-lasting bursts. On the right-hand panel the two 
	       theoretical distributions turn out to be indistinguishable.
             }
         \label{fig1569dwarfs}
   \end{figure}
%
%%%%%%%%%%%%%%%%%%%%%%%%%%%%%%%%%%%%%%%%%%%%%%%%

   Figs.~\ref{fig1569res}, \ref{fig1569moreres} and Table~\ref{tab1569res} 
   summarize and compare the results obtained with different models for 
   NGC\,1569 where the SFH and stellar yields are allowed to change one at a 
   time. The adopted structural parameters of the models are: 
   $M_{\mathrm{nor}}$~= 5~$\times$~10$^8$ M$_\odot$, $R_{\mathrm{eff}}$~= 1 
   kpc, $M_{\mathrm{d}}$~= 5~$\times$~10$^9$~M$_\odot$, 
   $R_{\mathrm{eff}}/R_{\mathrm{d}}$~= 0.1 (see Sect.~\ref{secModels} and 
   Table~\ref{tabcommonpar}). The infall law is exponentially decreasing with 
   time with $\tau$~= 0.5~Gyr for all the models (see, however, 
   Sect.~\ref{secMerge}). Models~\textsl{2aN} and \textsl{3aN} assume that a 
   long-lasting SF activity preceded the strong last 1--2 Gyr activity 
   detected with \textsl{HST}. Following Greggio et al. (1998) and Angeretti 
   et al. (2005), we assume that the most recent SF activity proceeded through 
   two violent successive bursts: the first started almost 2 
   (Model~\textsl{2aN}) to 1 (Model~\textsl{3aN}) Gyr ago and lasted nearly 2 
   (Model~\textsl{2aN}) to 1 (Model~\textsl{3aN}) Gyr, the second started 100 
   Myr ago and finished a few Myr ago. A 50 Myr gap separates the two 
   episodes; assuming a longer interburst period (150 Myr -- Angeretti et al. 
   2005) does not change the results. A weaker ($\nu$~= 0.05~Gyr$^{-1}$), but 
   long-lasting ($\Delta t$~= 6~Gyr) SF episode is assumed to have occurred 
   from $t$~= 5.55 to $t$~= 11.55~Gyr, where no detailed information is 
   available from \textsl{HST} photometry. This pristine SF raises the metal 
   content of the gas from $Z$~= 0 to $Z \simeq$~2~$\times$~10$^{-3}$. For 
   Models~\textsl{4aN} and \textsl{4aW}, the most ancient activity consists of 
   three intense short-lasting bursts, while Models~\textsl{5aN, 5aH, 5aM} 
   adopt a continuous, very low-level SFR over almost the whole Hubble time. 
   All the models share the same stellar nucleosynthesis prescriptions, but 
   Models~\textsl{4aW, 5aH, 5aM}. The adopted GW efficiencies are always the 
   same (see Table~\ref{tabpar}). 

   First, we want to emphasize that models assuming either a bursting, or a 
   gasping, or a fairly continuous, highly variable SF regime are all able 
   to produce results consistent with the observations (e.g., 
   Models~\textsl{2aN, 4aN, 5aN,} Fig.~\ref{fig1569res}, \emph{first, third} 
   and \emph{last col.,} respectively). They provide relatively small 
   differences in the predicted frequency distributions of low-mass stars with 
   metallicity (Fig.~\ref{fig1569dwarfs}). Only much deeper CMDs could unveil 
   the `true' SFH of NGC\,1569 at epochs prior to 1--2 Gyr ago. So far, we 
   have some clues that early SF in NGC\,1569 went on at rates significantly 
   lower than the ones characterizing later episodes (Angeretti et al. 2005). 
   This information, coupled to current observational evidence pointing to 
   fairly continuous SFHs for most DIGs and BCDs (Schulte-Ladbeck et al. 2001; 
   Annibali et al. 2003; Tosi 2003), leads us to indicate the SFH labelled 
   \textsl{5} as the most likely evolutionary scenario for NGC\,1569.

   Second, we find that for NGC\,1569 different combinations of stellar yields 
   reproduce the N/O data (at 2-$\sigma$, but fine tuning of model parameters 
   may reconcile model predictions to observations even better), except for 
   Model~\textsl{4aW}, adopting the van den Hoek \& Gronewegen (1997) yields 
   with metal-dependent $\eta_{\mathrm{AGB}}$ for LIMS. This model indeed 
   largely overestimates the N/O value measured in the local ISM of NGC\,1569 
   [log(N/O) = $-$1.39 $\pm$ 0.05; Kobulnicky \& Skillman 1997]. Successful 
   models for NGC\,1569 all have $\beta_{\mathrm{N}}/\beta_{\mathrm{O}}$ = 1, 
   at variance with models for NGC\,1705, where, in order to fit the 
   observations, a value of $\beta_{\mathrm{N}}/\beta_{\mathrm{O}}$ greater 
   than 1 is always required, independently of the choice of the stellar 
   yields.
   
   The issue of the GW efficiencies deserves further considerations. The SF 
   activity in NGC\,1569 triggers and sustains an outflow on a galactic scale 
   in our model galaxies. The more efficient is the SF process, the more 
   effective is the wind in removing the newly produced metals, thanks to its 
   dependence on the SN rate. For instance, Model~\textsl{3aN} 
   (Fig.~\ref{fig1569res}, \emph{second col.}), computed assuming a SF 
   efficiency higher than Model~\textsl{2aN} (Fig.~\ref{fig1569res}, 
   \emph{first col.}) during the last 1 Gyr, does not overproduce the 
   present-day oxygen content and overall metallicity of the gas even if the 
   $\beta_i$ values are kept the same, because the efficiency of gas removal 
   is increased in lockstep with the predicted higher Type II plus Type Ia SN 
   rate. In particular, we find that the assumed time modulation of the mass 
   loss rate [Equation~(\ref{eqWind})] leads to a predicted mass ejection rate 
   at the present time which is of the same order of magnitude of the 
   present-day SFR, in agreement with recent \textsl{Chandra} observations 
   (Martin et al. 2002; cfr. \emph{sixth} vs \emph{twelfth col.} in 
   Table~\ref{tab1569res}). This would not be the case if the mass loss rate 
   were simply proportional to the gas content. In that case, the rate of mass 
   loss through the wind would be maximum when the thermal energy equals the 
   binding energy of the gas for the first time, after which it would 
   monotonically decline in time, owing to the lower and lower gas amount left 
   over by the ongoing SF process. In other words, a simple dependence of the 
   outflow rate on the gas mass would not allow to predict enhanced mass loss 
   rates to coincide with enhanced SF episodes, as observed. It is also worth 
   emphasizing that by assuming a GW simply proportional to the SFR (as often 
   found in the literature) one forgets the effect of delayed SNIa explosions 
   and the wind goes to zero as soon as the SF stops.
 
   \subsubsection{The energetics of the ISM}
   \label{secEner}

   The energetics of the ISM is a complicated issue, and no recipes can be 
   found in the literature which treat it without introducing a number of 
   uncertain free parameters. The energy injected into the ISM by a typical 
   massive star through stellar winds during its lifetime is 
   $E_{\mathrm{wind}} \simeq$ 10$^{49}$ erg, to be compared with an initial 
   blast wave energy of $E_{\mathrm{SN}} \simeq$ 10$^{51}$ erg for both SN 
   types. What really matters is the fraction of this energy which is not 
   radiated away and remains effectively stored in the ISM. In previous work 
   on I\,Zw\,18, Recchi et al. (2001, 2004) assigned a low thermalization 
   efficiency of only 3 per cent to SNeII, whereas a much higher value, 100 
   per cent, was assumed for SNeIa, exploding in a medium already heated by 
   previous SNII activity after a non-negligible time lag from the onset of 
   the SF. However, those values are likely to apply strictly only to the case 
   of single, instantaneous bursts of SF. If the SF is an almost continuous 
   process, $\eta_{\mathrm{SNII}}$ must probably be increased (e.g. Strickland 
   \& Stevens 2000). Moreover, $\eta_{\mathrm{SNII}}$ should vary with time, 
   since it depends on the amount of gas in form of cold clouds at each time 
   (Melioli \& de Gouveia Dal Pino 2004). Since we do not follow the exchange 
   of matter between cold and hot gas phases in our models, we cannot state 
   how much gas is in the form of cold clouds at each time. Therefore, we 
   do not vary $\eta_{\mathrm{SNII}}$ and $\eta_{\mathrm{SNIa}}$ with time, 
   and assume average constant values for these parameters during the whole 
   galaxy evolution. Moreover, we do not treat SF in small (pc-sized) zones, 
   but we assume that the SF process is distributed over the whole physical 
   dimensions of the object, i.e. on kpc scale. Yet, our simplistic models 
   tell us something about the mechanisms regulating dwarf galaxy formation 
   and evolution. In the previous case of NGC\,1705, we have assigned the same 
   thermalization efficiency of 20 per cent to stellar winds, SNeII and SNeIa 
   (see Table~\ref{tabcommonpar}). Here, in the case of NGC\,1569, we discuss 
   results from models in which both $\eta_{\mathrm{SNII}}$ and 
   $\eta_{\mathrm{SNIa}}$ are changed. In two cases, also the prescriptions 
   about the DM content are modified. This affects the depth of the potential 
   well, and thus the time of the onset of the galactic outflow.

   In Table~\ref{tab1569eta} we list results from models analogous to 
   Model~\textsl{5aN} (Table~\ref{tab1569eta}, \emph{first row}) as far as the 
   prescriptions about the SFH, stellar nucleosynthesis and GW efficiencies 
   $\beta_i$ are concerned, but in which different values for 
   $\eta_{\mathrm{SNII}}$ and $\eta_{\mathrm{SNIa}}$ are adopted. In the 
   extreme case where $\eta_{\mathrm{SNII}}$ is set to 0.03 and 
   $\eta_{\mathrm{SNIa}}$ is set to 1, the time of occurrence of the GW is 
   delayed ($t_{\mathrm{GW}}~\simeq$ 4~Gyr rather than 2.5~Gyr; 
   Table~\ref{tab1569eta}, \emph{last col., first} vs. \emph{second row}). The 
   net outflow rate is reduced, since more weight is given to the less 
   numerous SNeIa, according to Equation~(\ref{eqWind}). As a result, we 
   predict a larger present-day gaseous mass, a higher ISM metallicity at the 
   present time, and a lower mass loss rate, which does not track the SFR any 
   more. This is due to the fact that, while the SNII rate closely follows the 
   SFR, the SNIa rate displays a very different trend. Therefore, while the 
   energy injection from SNeII proceeds in lockstep with the SFR, the bulk of 
   the explosion energy from SNeIa is restored to the ISM with a delay time 
   which depends on the properties of the SF burst. For Model~\textsl{5aN}, 
   this delay time is $\sim$~2~Gyr for the early weaker activity and 
   $\sim$~1~Gyr for the second to last burst. SNIa progenitors born during the 
   most recent SF activity had no time to restore the bulk of their energy 
   into the ISM yet. If one increases the values of the $\beta_i$ parameters 
   -- which are free parameters of the model -- in order to reproduce the 
   observational constraints, finds that by assuming higher $\beta_i$ values 
   ($\sim$~5~$\times$~10$^{-6}$ for H, D and He and 
   $\sim$~5~$\times$~10$^{-5}$ for the elements heavier than helium) can fit 
   all the available observational constraints, but the present-day mass loss 
   rate, which turns out to be roughly a factor of three lower than the 
   observed present-day SFR. This seems to suggest that $\eta_{\mathrm{SNII}} 
   \simeq \eta_{\mathrm{SNIa}}$ is a more reasonable choice.

   In Table~\ref{tab1569eta} we also show the results obtained in a case where 
   $\eta_{\mathrm{SNII}} = \eta_{\mathrm{SNIa}}$ = 0.50. In this case, a lower 
   present-day gaseous mass survives the SF process \emph{(fourth col., third 
   row)}, unless one reduces the strength of the $\beta_i$ coefficients. It is 
   very difficult to put constraints on the values of the $\beta_i$ and 
   $\eta_{\mathrm{SNII}}$, $\eta_{\mathrm{SNIa}}$ parameters separately. In 
   fact, they enter the model in a complex way, i.e. through the products 
   $\beta_i \times \eta_{\mathrm{SNII}}$, $\beta_i \times 
   \eta_{\mathrm{SNIa}}$, which are then used to weigh the SN rates. The 
   poorly known DM amount and distribution do not seem to cause further 
   complications: in Table~\ref{tab1569eta} we display the results obtained by 
   a model in which $M_{\mathrm{d}}/M_{\mathrm{lum}}$ is set to 50 rather than 
   10 and $\eta_{\mathrm{SNII}} = \eta_{\mathrm{SNIa}}$ = 0.50 \emph{(fourth 
   row)}. It can be seen that, although in this case $t_{\mathrm{GW}} \simeq$ 
   2 rather than $\sim$~1 Gyr, the model predictions are almost the same (cfr. 
   \emph{third} vs. \emph{fourth rows} in Table~\ref{tab1569eta}), due to the 
   scarce evolution of the system at those early times. Similarly, setting 
   $M_{\mathrm{d}}/M_{\mathrm{lum}}$ to 5 and $\eta_{\mathrm{SNII}} = 
   \eta_{\mathrm{SNIa}}$ = 0.20 does not affect the model predictions (Table 
   7, \emph{last row}).

   \subsubsection{Late accretion of gaseous lumps onto NGC\,1569?}
   \label{secMerge}

%%%%%%%%%%%%%%%%%%%%%%%%%%%%%%%%%%%%%%%%%%%%%%%% PLEASE, TWO COLUMNS FIGURE!!!
%
   \begin{figure*}
   \psfig{figure=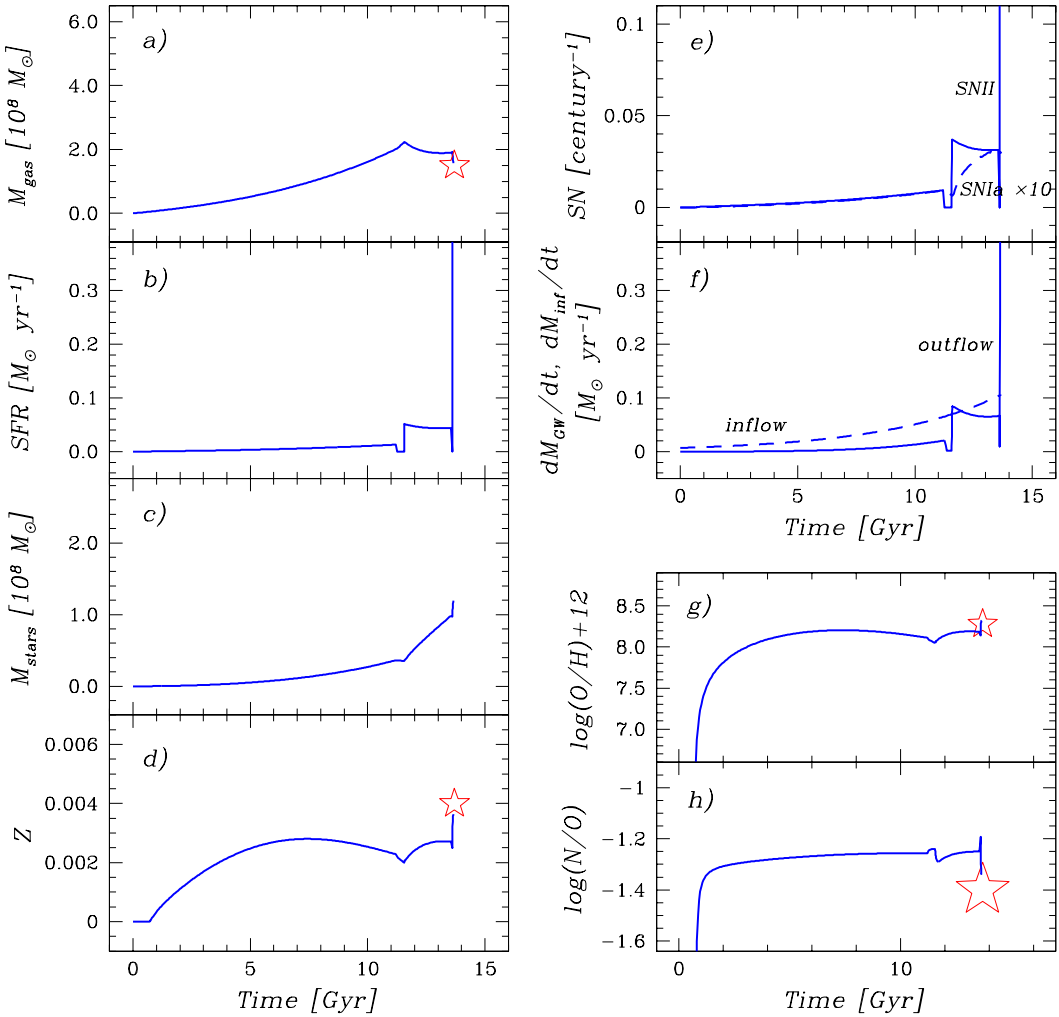,width=15cm}
      \caption{ Predictions of Model~\textsl{5dN} for NGC\,1569: a) 
	        $M_{\mathrm{gas}}$, b) SFR, c) $M_{\mathrm{stars}}$, d) gas 
	        metallicity, e) Type II \emph{(solid line)} and Type Ia 
		\emph{(dashed line;} multiplied by a factor of 10 to make it 
		clearly visible) SN rate, f) $\dot{M}_{\mathrm{GW}}$ 
		\emph{(solid line)} and $\dot{M}_{\mathrm{inf}}$ \emph{(dashed 
		line)} as functions of time; g) log(O/H)+12 and h) log(N/O) as 
		functions of time. The stars represent observed present-time 
		values (see text for details).
             }
         \label{fig15695dN}
   \end{figure*}
%
%%%%%%%%%%%%%%%%%%%%%%%%%%%%%%%%%%%%%%%%%%%%%%%%

   It has been suggested (Melioli \& de Gouveia Dal Pino 2004) that infall of 
   fresh gas from the surroundings could help to keep the value of the 
   thermalization efficiencies lower than unity. Therefore, we have computed a 
   model (Model~\textsl{5dN}) in which the infall rate increases with time and 
   the time scale for infall, $\tau$, is set to 5 rather than 0.5 Gyr, 
   simulating a late accretion of gaseous lumps. With this choice, the 
   present-day properties of NGC\,1569 are well reproduced (see 
   Fig.~\ref{fig15695dN}). The unusually high SFR in NGC\,1569 is now 
   explained by the replenishment of the H\,{\small I} reservoir in its disc 
   by newly accreted material. Indeed, some evidence that the galaxy is 
   presently ingesting a small gaseous companion exists (Stil \& Israel 1998; 
   M\"uhle et al. 2005). In particular, according to M\"uhle et al., the 
   emergence of the latest burst could be explained by the accretion of 
   H\,{\small I} gas with a mass of $\sim$ 10$^7$ M$_\odot$. Our 
   Model~\textsl{5dN} is characterized by an infall rate $\dot 
   M_{\mathrm{inf}} \sim$ 0.1 M$_\odot$ yr$^{-1}$ during the past 100 Myr, 
   which implies that during this period nearly 10$^7$ M$_\odot$ of primordial 
   gas are accreted by the galaxy. The predicted outflow rate is $\dot 
   M_{\mathrm{GW}} \sim$ 0.6 M$_\odot$ yr$^{-1}$, which leads to a total 
   ejected mass of 6.3 $\times$ 10$^6$ M$_\odot$ over the last 10 Myr, to be 
   compared with an upper limit to the wind mass of 6.2 $\times$ 10$^6$ 
   M$_\odot$ obtained from spectral fitting (Martin et al. 2002).

   \subsection{The problem of nitrogen: NGC\,1569 vs NGC\,1705} 
   \label{secN}

   Understanding the evolution of nitrogen is one of the major challenges of 
   modern astrophysics. The situation is far from being settled even in our 
   own Galaxy. Recent abundance data for very metal-poor Galactic halo stars 
   (Israelian et al. 2004; Spite et al. 2005) suggest that an important 
   production of primary N took place in the massive halo stars, while the 
   delayed N production from intermediate-mass stars likely overwhelms any 
   massive star contribution at later times (e.g. Chiappini, Romano \& 
   Matteucci 2003a). Yet, current massive star models do not produce enough 
   nitrogen to fit the halo data (see Chiappini, Matteucci \& Ballero 2005 for 
   an extensive discussion).

   Here we find that, while the N/O ratio in the local ISM of NGC\,1569 is 
   rather well reproduced by models assuming 
   $\beta_{\mathrm{N}}/\beta_{\mathrm{O}}$~= 1, in order to explain the low 
   N/O ratio measured in H\,{\small II} regions belonging to NGC\,1705, it is 
   necessary to assume $\beta_{\mathrm{N}}/\beta_{\mathrm{O}}~\neq$ 1. This 
   result is independent of the adopted stellar yields and infall law. 
   Furthermore, when steepening the IMF slope in NGC\,1705 as suggested by 
   Annibali et al. (2003; $x$~= 1.6 rather than 1.35), the discrepancy between 
   model predictions and observations increases. Before concluding that GWs in 
   NGC\,1705 must eject nitrogen more effectively than in NGC\,1569, there is 
   one more possibility we have to examine.

   \subsubsection{IMF variations from galaxy to galaxy?} \label{secresIMF}

   One may wonder whether coupling an IMF different from Salpeter in NGC\,1569 
   with the same $\beta_{\mathrm{N}}/\beta_{\mathrm{O}}$ value which matches 
   the N/O data for NGC\,1705 would allow to reproduce the N/O ratio measured 
   for NGC\,1569 as well. Possible IMF variations in the range of massive and 
   intermediate-mass stars are constrained by \textsl{HST} photometry. In 
   particular, for NGC\,1569 a power-law exponent flatter than Salpeter's 
   cannot be excluded. Indeed, a value of $x$ = 0.35 over the whole stellar 
   mass range still reproduces the red plume of the observed CMD (Angeretti et 
   al. 2005), though it should be cautioned that, due to NICMOS resolution 
   limits, small unresolved stellar clusters might appear as single stars 
   populating the red plume, thus leading to an artificial flattening of the 
   IMF. Besides this, one should always bear in mind that the observed CMDs 
   cannot constrain the IMF below the mass limit given by the galaxy distance 
   and the photometric depth of the images.

   In the following, we show results obtained by running modified versions of 
   Model~\textsl{5aN}, which assume different IMFs, still fullfilling the 
   required photometric properties of NGC\,1569. We adopt either $x$ = 0.35 
   over the whole stellar mass range, or a two-slope IMF consistent with 
   Salpeter down to 0.5 M$_\odot$, and then flattening to $x$ = 0.3 for 0.1 
   $\le m/$M$_\odot \le$ 0.5 (Kroupa \& Weidner 2005).

%%%%%%%%%%%%%%%%%%%%%%%%%%%%%%%%%%%%%%%%%%%%%%%%
%
   \begin{figure}
   \psfig{figure=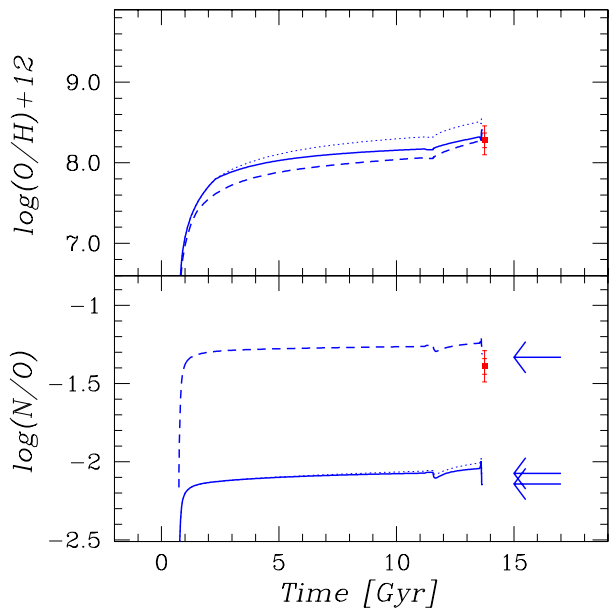,width=\columnwidth}
      \caption{ Temporal evolution of oxygen \emph{(top panel)} and nitrogen 
               to oxygen \emph{(bottom panel)} in the ISM of NGC\,1569. Model 
               predictions for different IMF slopes [$x$ = 1.35 (0.30) above 
               (below) 0.5 M$_\odot$, \emph{dashed lines}; $x$ = 0.35 for the 
               whole stellar mass range, \emph{dotted} and \emph{solid 
               lines}) and metal ejection efficiencies ($\beta_i$ = 2 $\times$ 
	       10$^{-5}$, \emph{dashed} and \emph{dotted lines}; $\beta_i 
	       \simeq$ 3 $\times$ 10$^{-5}$, \emph{solid lines}) are compared 
	       to the available observations (\emph{filled squares;} see 
	       text). The ending points of the evolutionary tracks are marked 
	       with arrows in the log(N/O) vs. time plot.
              }
         \label{fig1569no2}
   \end{figure}
%
%%%%%%%%%%%%%%%%%%%%%%%%%%%%%%%%%%%%%%%%%%%%%%%%

   The input SFRs are modified accordingly (a straightforward scaling law does 
   not apply -- see Angeretti et al. 2005, their figure 6). Extrapolating the 
   SFR below 0.5~M$_\odot$ by adopting $x$~= 0.3 rather than $x$~= 1.35, 
   results in a SFR about 30 per cent lower than assumed up to now (Angeretti 
   et al. 2005). As a consequence, a lower final mass in stars of 
   $M_{\mathrm{stars}}(t_{\mathrm{now}})~\sim$ 8~$\times$~10$^7$~M$_\odot$ 
   and a higher present-day gaseous mass of 
   $M_{\mathrm{gas}}(t_{\mathrm{now}})~\sim$ 2.5~$\times$~10$^8$~M$_\odot$ 
   are predicted by Model~\textsl{5aN}. The predicted current metallicity of 
   the galaxy is $Z$ = 0.003, while the theoretical N/O ratio is almost 
   unchanged, log(N/O)~= $-$1.33 at the present time 
   (cfr. Fig.~\ref{fig1569res}, \emph{last col., last row} to 
   Fig.~\ref{fig1569no2}, \emph{bottom panel, dashed line}). However, since 
   the SF preceding the last 1--2 Gyr activity unraveled by \textsl{HST} is a 
   poorly-constrained parameter of the model, we can enhance it with respect 
   to the prescriptions of Model~\textsl{5aN}. For example, by setting $\nu$~= 
   0.02~Gyr$^{-1}$ rather than 0.012~Gyr$^{-1}$, we find a final stellar 
   (gaseous) mass of 1~$\times$~10$^8$~M$_\odot$ 
   (1.8~$\times$~10$^8$~M$_\odot$). The final metallicity of the ISM is $Z$~= 
   0.004. Therefore, a two-slope IMF consistent with Salpeter's one above 
   0.5~M$_\odot$ and flattening below ($x$~= 0.3) is permitted by the models.

   If we assume instead $x$~= 0.35 over the whole stellar mass range, we find 
   a stellar (gaseous) mass of 6~$\times$~10$^6$~M$_\odot$ 
   (2.9~$\times$~10$^8$ M$_\odot$), whereas a metallicity as high as $Z$~= 
   0.006 is attained, so the model cannot be accepted. Notice that by 
   increasing the metal ejection efficiency (i.e., the $\beta_i$ parameter 
   value) one can recover the present-day metallicity, but the predicted 
   nitrogen to oxygen abundance ratio is always out of the range allowed by 
   the observations. This can be clearly seen from Fig.~\ref{fig1569no2}, 
   where the \emph{dotted lines} stand for a model analogous to 
   Model~\textsl{5aN}, in which the IMF slope is set to 0.35 rather than 1.35, 
   while the \emph{solid lines} refer to the same model once the metal 
   ejection efficiency is increased in such a way that $Z$~= 0.004 at the 
   present time.

   In conclusion: (i) IMF variations compatible with the shape of the CMD do 
   not prove useful in solving the puzzle of the different log(N/O) measured 
   for NGC\,1569 and NGC\,1705; (ii) flattening the IMF for 0.1~$< m$/M$_\odot 
   <$ 0.5 in NGC\,1569 produces results which still agree with the 
   observations, whereas (iii) adopting a flat IMF over the whole stellar mass 
   range produces results at variance with the observational data. This may be 
   a hint that small unresolved stellar clusters have really been treated as 
   single stars located on the red plume in the CMD analysis. We thus conclude 
   that \emph{different efficiencies of nitrogen ejection through the winds 
   are needed in order to explain both the NGC\,1569 and NGC\,1705 relevant 
   data.}

   \section{Discussion and future work}

\begin{figure}
   \psfig{figure=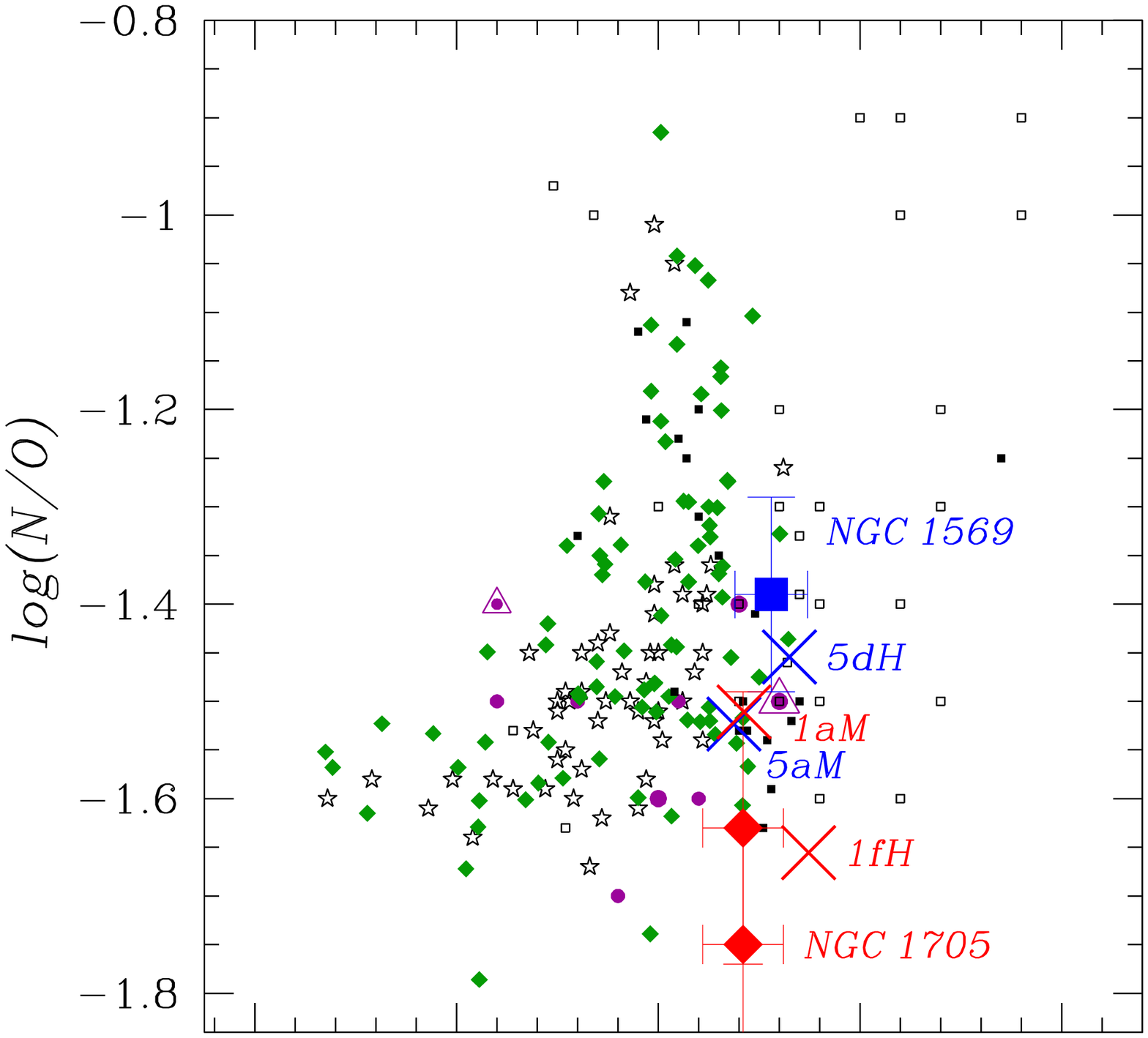,width=\columnwidth}
   \psfig{figure=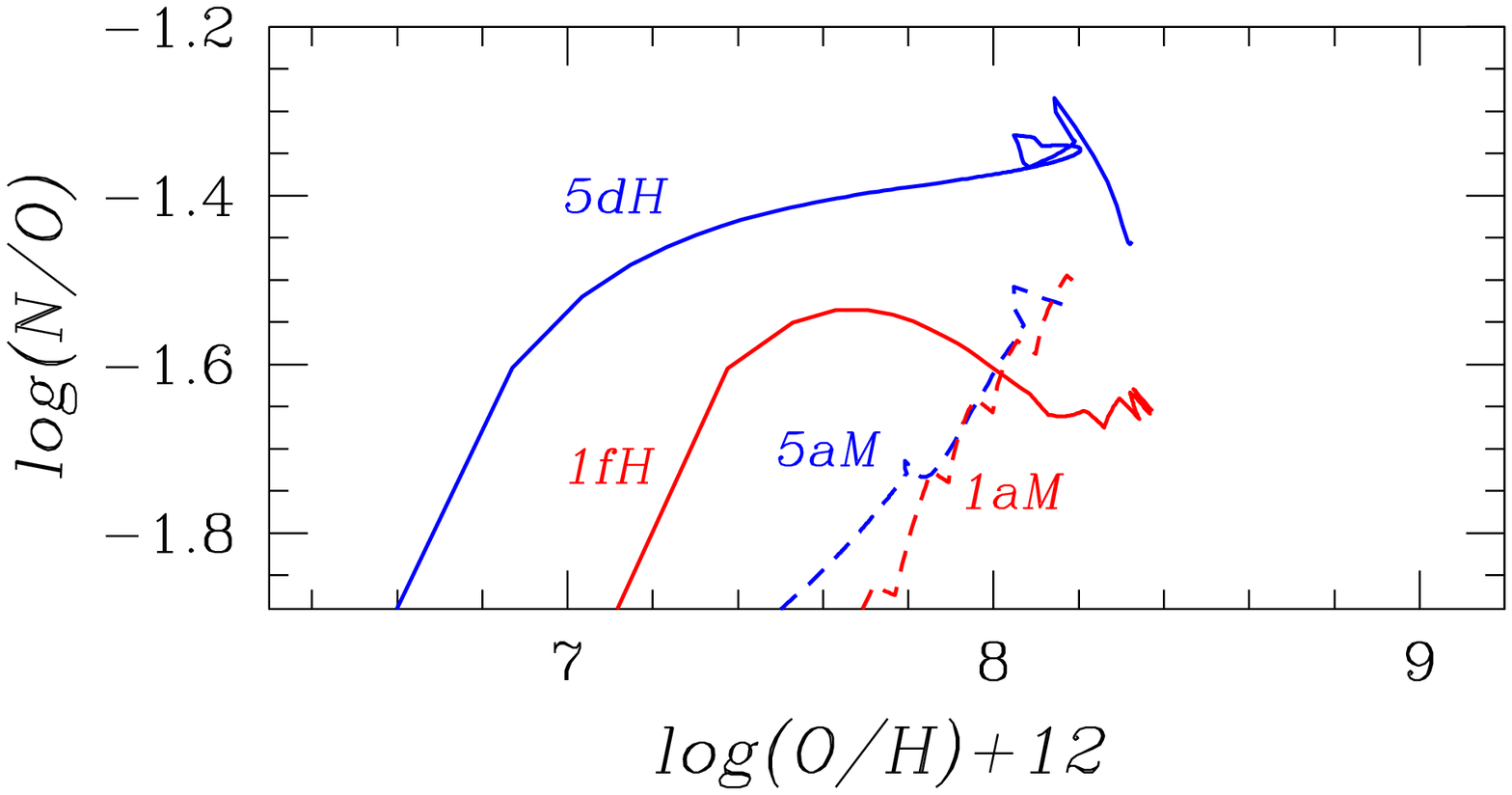,width=\columnwidth}
      \caption{ \emph{Upper panel:} log(N/O) vs. log(O/H)+12 diagram for 
	        several samples of DIGs, BCDs, and low surface brightness 
		dwarf galaxies. \emph{Filled diamonds:} metal-poor H\,{\tiny 
                II} galaxies from Kobulnicky \& Skillman (1996); \emph{filled 
		circles:} low surface brightness dwarfs from van Zee et al. 
		(1997): bigger circles are objects with higher average SFR in 
		the past; the big empty triangle stands for the sample object 
		with the highest SFR at present -- 0.35 M$_\odot$ yr$^{-1}$, 
		while the small empty one stands for the object with the 
		lowest SFR at present -- 0.0015 M$_\odot$ yr$^{-1}$; 
		\emph{stars:} ground-based spectroscopic observations of 54 
		supergiant H\,{\tiny II} regions in 50 low-metallicity BCDs 
		from Izotov \& Thuan (1999); \emph{squares:} Virgo DIGs and 
		BCDs from Lee et al. (2003b; \emph{small filled squares}) and 
		V\'\i lchez \& Iglesias-P\'aramo (2003; \emph{small empty 
		squares}). The big diamonds and the big square on the right 
		represent measurements for NGC\,1705 and NGC\,1569, 
		respectively, along with their 2-$\sigma$ errors. Also shown 
		are predictions from Models~\textsl{1aM}, \textsl{5aM}, 
		\textsl{1fH} and \textsl{5dH} (final points of the tracks; 
		\emph{big crosses}). A higher efficiency of nitrogen ejection 
		through the wind is assumed for Model~\textsl{1fH} with 
		respect to Models~\textsl{1aM}, \textsl{5aM} and \textsl{5dH} 
		(see text for details). \emph{Lower panel:} full evolutionary 
		tracks for Models~\textsl{1aM}, \textsl{5aM}, \textsl{1fH} and 
		\textsl{5dH} in the log(N/O)--log(O/H)+12 space.
              }
         \label{figAll}
   \end{figure}

   In this section we extend the previous discussion to the interpretation of 
   the log(N/O) vs. log(O/H)+12 diagram for a wider sample of DIGs and BCDs. 
   In Fig.~\ref{figAll}, \emph{upper panel,} we show galaxies for which both 
   oxygen and nitrogen have been measured. The position of NGC\,1705 and 
   NGC\,1569 in the diagram is given by the big filled diamonds\footnote{For 
   NGC\,1705, both the mean N/O from measurements in all H\,{\tiny II} regions 
   and that without the extremely low nitrogen abundance of region B4 are 
   considered (see Lee \& Skillman 2004).} and the big filled square with 
   2-$\sigma$ error bars, respectively (see Sect.~\ref{secStarFor} for 
   references). We remind the reader that this is not an evolutionary diagram: 
   it refers to present-time abundances, derived from H\,{\small II} region 
   spectroscopy.

   Several features emerge from this diagram. It has been pointed out 
   (Izotov \& Thuan 1999) that the N/O ratio is independent of oxygen for BCDs 
   with log(O/H)+12 $\le$ 7.6. As already discussed by Matteucci \& Tosi 
   (1985), this is easily interpreted if the fraction of primary N is at least 
   30 per cent. Izotov \& Thuan interpret this result as an indication for 
   primary nitrogen being produced by the same massive stars which synthesize 
   oxygen. In their view, the small dispersion in N/O at these low 
   metallicities argues against any delayed primary nitrogen production from 
   intermediate-mass stars, thus suggesting that these galaxies are now 
   undergoing their first burst of SF. The delayed release of primary N would 
   not leave its imprints until metallicities 7.6 $<$ log(O/H)+12 $<$ 8.2 are 
   attained, while secondary N production would become important only when 
   galaxies get enriched to log(O/H)+12 $\ge$ 8.2 (see also Henry, Edmunds \& 
   K\"oppen 2000). These mechanisms would be responsible for the scatter in 
   N/O observed at log(O/H)+12 $\ge$ 7.6.

%%%%%%%%%%%%%%%%%%%%%%%%%%%%%%%%%%%%%%%%%%%%%%%%
%
%
%%%%%%%%%%%%%%%%%%%%%%%%%%%%%%%%%%%%%%%%%%%%%%%%

   In previous works (Matteucci \& Tosi 1985; Marconi et al. 1994; Chiappini 
   et al. 2003a) we were able to populate the observed log(O/H)+12--log(N/O) 
   diagram with model galaxies, without invoking a primary N production from 
   massive stars (see also Henry et al. 2000). We found that dwarf galaxies 
   experiencing an initial burst of SF in the past and a more recent one, or, 
   alternatively, just one SF episode lasting long enough to let 
   intermediate-mass stars have time restoring their nucleosynthesis products, 
   tend to cluster around the N/O `plateau'. Models in which more than one 
   burst occur populate the high metallicity, high N/O part of the plot (e.g., 
   Chiappini et al. 2003a, their figure 13 and table 5). However, in those 
   models the SFH and IMF were treated as free parameters, a simpler 
   dependence of the outflow rate on the SFR was assumed, and no attempts were 
   made at modelling specific objects.

   As far as an important production of primary N from very metal-poor massive 
   stars is concerned, we recall that current massive star models do not 
   predict an amount of primary N sufficient to explain the Galactic halo star 
   data (Chiappini et al. 2005, and references therein). On the 
   other hand, the amount of primary N synthesized in low-metallicity 
   intermediate-mass stars due to the third dredge-up and HBB processes has 
   been probably overestimated (e.g. Chiappini et al. 2003a). Besides the 
   observational data, in Fig.~\ref{figAll} we plot also the predictions from 
   Models~\textsl{1aM} and \textsl{1fH} for NGC\,1705, and \textsl{5aM} and 
   \textsl{5dH} for NGC\,1569. In the upper panel, only the ending points of 
   the theoretical tracks are shown \emph{(big crosses)}, whereas in the lower 
   panel the complete evolutive tracks are displayed. NGC\,1705 and NGC\,1569 
   show very high SFRs -- among the highest ever observed for DIGs and BCDs -- 
   and evidence for protracted SFHs in the past. Therefore, it is not 
   surprising that they reach a relatively large metallicity. However, 
   NGC\,1705 has the lowest N/O ratio observed at those metallicities, and the 
   N/O ratio of NGC\,1569 is quite low too, in contrast with previous ideas 
   that DIGs and BCDs evolved through several bursts of SF or experiencing a 
   long-lasting SF regime with a high SF efficiency should populate the high 
   N/O region of the log(O/H)+12--log(N/O) plot. Models~\textsl{1fH} and 
   \textsl{5dH} have been computed with the van den Hoek \& Groenewegen (1997) 
   plus Nomoto et al. (1997) yields, for the case of minimum HBB extent. Both 
   assume an infall law exponentially increasing in time. The adopted stellar 
   IMF (Salpeter 1955) and SFHs are consistent with CMD analyses (see previous 
   sections). The case with minimum HBB extent well matches the log(N/O) value 
   actually observed for NGC\,1569. However, in order to make the model 
   predictions agree with the observations for NGC\,1705 too, one must require 
   that some other mechanism(s) is (are) at work. We argue 
   (Sect.~\ref{secRes1705}) that differential galactic outflows in which 
   different fractions of nitrogen are lost with respect to oxygen offer a 
   viable solution. For this reason, Model~\textsl{1fH} has been computed by 
   assuming $\beta_{\mathrm{N}}/\beta_{\mathrm{O}} \simeq$ 2 rather than 1, 
   namely, postulating a higher nitrogen ejection efficiency in the outflow of 
   NGC\,1705. Alternatively, in NGC\,1705 we might be seeing localized 
   self-pollution from newly-born massive stars. However, the constancy of the 
   oxygen abundance measured in several H\,{\small II} regions of NGC\,1705 
   poses serious challenges to this interpretation (see Sect.~\ref{parSelf}). 
   Using the yields from rotating star models by Meynet \& Maeder (2002) also 
   allows us to fit the data (Fig.~\ref{figAll}; Models~\textsl{1aM} and 
   \textsl{5aM}). However, in this case only the highest N/O value for 
   NGC\,1705 and the lowest one for NGC\,1569 can be explained, rather than 
   \emph{the difference} between the two. When faced with the whole data set, 
   we think that the observed scatter of log(N/O) values at relatively high 
   metallicities needs the hypothesis of differentially enriched GWs in order 
   to be explained. Clearly, only detailed study of a larger number of 
   individual dwarfs, for which the SFH and the IMF had been previously 
   inferred by means of CMD analyses, would allow us to understand the driving 
   processes which lead to the scatter. Recently, our group (PI: A. Aloisi) 
   has been granted time to perform \textsl{HST} observations of NGC\,4449, a 
   starburst DIG located at a distance $D \simeq$ 4.2 Mpc with log(O/H)+12~= 
   8.3 and log(N/O)~= $-$1.2 (Sabbadin, Ortolani \& Bianchini 1984), i.e., 
   with a N/O ratio higher than both NGC\,1569 and NGC\,1705. The use of the 
   Advanced Camera for Surveys (ACS) on board \textsl{HST} will allow us to 
   reach stars quite fainter than the red giant branch tip (if they exist), 
   and to infer the SFH of NGC\,4449 up to several Gyrs ago. We want also to 
   call attention to II\,Zw\,40, a galaxy located at a distance $D \sim$ 7.5 
   Mpc -- i.e., within reach of \textsl{HST} instrumentation -- with 
   log(O/H)+12~= 8.091$^{+0.013}_{-0.014}$, log(N/O) = 
   $-$1.052$^{+0.059}_{-0.077}$, He/H~= 0.093 $\pm$ 0.013 (Kobulnicky \& 
   Skillman 1996), and an old underlying stellar population (Vanzi et al. 
   1996). Given its very high N/O abundance ratio, it would be an interesting 
   object to study in the context of our research.

%   Finally, we want to stress that by assuming some primary N production 
%   through HBB in intermediate-mass stars, we obtain a mildly varying N/O 
%   ratio as a function of O/H over many Gyr, if a low-level, long-lasting SF 
%   regime is assumed (Fig.~\ref{figAll}, Model~\textsl{5dH}). In an evolutive 
%   scenario where a gasping rather than a bursting SF regime better 
%   characterizes both BCDs and DIGs, this primary N production would thus 
%   naturally tend to produce a plateau in N/O. However, we caution that the 
%   extent of HBB in low-metallicity environments is still matter of debate. 
%   This is an important point, since by lowering (increasing) the HBB 
%   efficiency one ends up with a lower (higher) N/O plateau level. 
%   Flattening the IMF shifts the N/O plateau as well: an extreme case is shown 
%   in Fig.~\ref{fig1569no2}. Finally, it is worth emphasizing that the 
%   important r\^ole of inflows and outflows of gas in galaxy formation has not 
%   been fully addressed yet, and their interplay still waits for being better 
%   understood.

   \section{Conclusions}

   In this paper we have presented one-zone chemical evolution models for two 
   dwarf starburst galaxies, NGC\,1705 and NGC\,1569. For both objects, we 
   have adopted the SFH and IMF previously inferred from \textsl{HST} optical 
   and NIR CMDs and found that strong galactic winds develop owing to their 
   violent SF activity. Our main conclusions can be summarized as follows:
   \begin{enumerate}
     \item The current metallicity of H\,{\small II} regions of both NGC\,1705 
           and NGC\,1569 is well reproduced only if strong GWs efficiently 
	   remove the metals from the galaxies, in agreement with previous 
	   works (e.g., Matteucci \& Tosi 1985; Pilyugin 1993; Carigi et al. 
	   1995; Recchi et al. 2001, 2004, 2005).
     \item The very low N/O ratio measured in NGC\,1705 (Lee \& Skillman 2004) 
           requires a higher efficiency of nitrogen ejection in the wind of 
	   NGC\,1705 relative to the wind of NGC\,1569 in order to be 
	   explained, a result independent of the other parameters of the 
	   model. Alternative solutions, such as complete suppression of HBB 
	   in intermediate-mass stars or localized chemical pollution from 
	   very massive stars in NGC\,1705, are discussed and seem unlikely.
     \item An offset exists between the oxygen abundance of the neutral phase 
           and that of the ionized gas in NGC\,1705, which seems absent in the 
           N/O data (Aloisi et al. 2005). Once some primary nitrogen 
           production through HBB in intermediate-mass stars is allowed for in 
           NGC\,1705, an almost constant N/O abundance ratio is predicted in 
	   its ISM during almost all the evolution, whereas the oxygen 
	   abundance always increases with time (see, e.g., 
	   Fig.~\ref{fig1705yields}, \emph{first} to \emph{third cols.}). This 
	   qualitatively explains why an observational offset should be 
	   present in O/H, and absent in N/O, though uncertainties in both the 
	   physics of the models and the treatment of the data prevent us from 
	   attempting a more quantitative analysis. Notice that by using the 
	   yields of Meynet \& Maeder (2002), computed by taking into account 
	   mass loss and stellar rotation, but neglecting HBB, the theoretical 
	   N/O ratio in NGC\,1705 steeply increases during its whole 
	   evolution (Fig.~\ref{fig1705yields}, \emph{last col., last row}). 
	   Therefore, in this case the constancy of the N/O ratio in the 
	   neutral and ionized gas phases cannot be explained.
     \item We find hints that \emph{the third dredge-up and HBB must be less 
           effective than assumed up to now in low-metallicity 
           intermediate-mass stars.} We do not find any need for a significant 
	   primary nitrogen production from massive stars to explain the 
	   present-day abundances in NGC\,1705 and NGC\,1569. The large 
	   primary nitrogen production which characterizes massive stars at 
	   $Z_{\mathrm{ini}} \le$ 10$^{-5}$ (see Chiappini et al. 2005) does 
	   not leave any imprint on the present-day properties of NGC\,1705 
	   and NGC\,1569, at $Z \sim$ 0.004.
     \item We demonstrate that the efficiency of hydrogen entrainment in the 
           outflow crucially depends on the adopted infall law. If we allow 
	   for important mergers with gaseous lumps at late times, we find 
	   that the efficiency of hydrogen ejection through the outflow must 
	   be of the same order of magnitude as that for the heavy elements in 
	   order to fit the observations. On the contrary, with an infall rate 
	   exponentially decreasing with time (as usually assumed by GCE 
	   models), the required efficiency of hydrogen ejection is up to 
	   one order of magnitude lower than that for the heavy elements. 
	   While some evidence exists that NGC\,1569 could be currently 
	   accreting a small companion cloud (Stil \& Israel 1998; M\"uhle et 
	   al. 2005), no clear signs of external triggers to the sudden, 
	   intense ongoing burst observed in NGC\,1705 have been found up to 
	   now (Meurer et al. 1998).
     \item By assuming that the GW efficiency varies in lockstep with the 
           predicted Type II plus Type Ia SN rate, we find 
	   $\dot{M}_{\mathrm{GW}}(t_{\mathrm{now}}) \sim 
	   \psi(t_{\mathrm{now}})$ in NGC\,1569 [where 
	   $\dot{M}_{\mathrm{GW}}(t_{\mathrm{now}})$ and 
	   $\psi(t_{\mathrm{now}})$ are the current mass loss and SF rates], 
	   independently of the assumed infall law and in agreement with 
	   observations. The total mass of gas ejected at the average rate 
	   predited during the last 10 Myr is 6.3~$\times$~10$^6$~M$_\odot$ in 
	   the case of an infall rate exponentially increasing with time, 
	   while it is 3.5~$\times$~10$^6$~M$_\odot$ in the case of an 
	   exponentially decreasing infall. These values agree with the 
	   best-fitting value (3.5~$\times$~10$^6$~M$_\odot$) and upper limit 
	   (6.2~$\times$~10$^6$~M$_\odot$) for the mass ejected in the outflow 
	   given by Martin et al. (2002). Current estimates of the gas mass 
	   in the outflow of NGC\,1705 vary by almost three orders of 
	   magnitude, going from 
	   $\dot{M}_{\mathrm{GW}}(t_{\mathrm{now}})~\simeq$ 
	   0.0026~M$_\odot$~yr$^{-1}$ up to $\sim$~5~M$_\odot$~yr$^{-1}$ 
	   (Meurer et al. 1992). These values do not usefully constrain our 
	   models, which predict mass loss rates ranging from 
	   $\dot{M}_{\mathrm{GW}}(t_{\mathrm{now}})~\simeq$ 0.04 to 
	   0.2~M$_\odot$~yr$^{-1}$, depending on the adopted infall law 
	   prescriptions.
   \end{enumerate}
   Clearly, we need to apply our models to a larger galaxy sample before 
   drawing firmer conclusions about late-type dwarf galaxy formation and 
   evolution. We identify NGC\,4449 and II\,Zw\,40 as promising candidates: 
   owing to their metallicities, similar to those of NGC\,1705 and NGC\,1569, 
   but higher N/O ratios, they probe a different locus in the log(N/O) vs. 
   log(O/H)+12 plane. Thus, they could help us to unravel the origin of the 
   observed spread among N/O values at a given metallicity. In particular, our 
   group has already been granted \textsl{HST} time to perform ACS 
   observations of NGC\,4449. These will enable us to determine its SFH and 
   IMF, a necessary step to work out the appropriate chemical evolution models.

   \section*{Acknowledgments}

   We are grateful to A. Aloisi, D. Calzetti, A. D'Ercole, H. Lee, S. Recchi, 
   R. Sancisi, E. Skillman and D. Strickland for illuminating discussions. DR 
   and MT acknowledge useful discussions with the members of the LoLa-GE Team 
   at the International Space Science Institute (ISSI) in Bern. We are 
   indebted to the anonymous referee, whose constructive criticism led us to 
   analyse more thoroughly the parameter space of the models. Financial 
   support from INAF through project `Blue Compact Galaxies: Primordial Helium 
   and Chemical Evolution' is also acknowledged, as well as financial support 
   from MIUR through project COFIN 2003 prot. n.~028039 `Chemical and 
   Dynamical Evolution of Galaxies: Interpretation of Abundance Patterns in 
   the Universe'.

\bsp

\label{lastpage}

\end{document}